\documentclass{naturep}
\usepackage[utf8]{inputenc}
\usepackage{lineno}
\usepackage{setspace}
\usepackage{color}
\usepackage{graphicx}
\usepackage{booktabs}
\usepackage{amssymb}
\usepackage{amsmath}
\usepackage{hyperref}
\usepackage{adjustbox}
\usepackage{subcaption}
\usepackage{tablefootnote}
\usepackage[margin=0.6in]{geometry}
\usepackage{verbatim}
\usepackage{caption}
\usepackage{paralist}
\usepackage{cleveref}
\usepackage{multirow}
\usepackage{tabularx} 
\usepackage{array} 

\usepackage{multirow}%
\usepackage{amsmath,amssymb,amsfonts}%
\usepackage{amsthm}%
\usepackage{mathrsfs}%
\usepackage[title]{appendix}%
\usepackage{xcolor}%
\usepackage{textcomp}%
\usepackage{manyfoot}%
\usepackage{booktabs}%
\usepackage{algorithm}%
\usepackage{algorithmicx}%
\usepackage{algpseudocode}%
\usepackage{listings}%
\usepackage{booktabs}
\usepackage{float}
\usepackage{microtype}
\usepackage{xurl}

\newcommand\Heading[1]{
  \noindent\textbf{\Large{#1}}
}

\newcommand\heading[1]{
  \noindent\textbf{\large{#1}}
}

\title{\raggedright{
SpCAST enables scalable and interpretable integration of single-cell RNA sequencing and single-cell-resolved spatial transcriptomics
}}

\author{
Yiyang Zhang$^{1,5,6,\ast}$,
Bokai Zhao$^{2,5,6,\ast}$,
Xiaoru Zhang$^{1,5,6}$,
Zongchang Du$^{8}$,
Minfeng Xu$^{10}$,
Xiaojuan Sun$^{1,9,\dag}$,
Tianzi Jiang$^{3,4,5,6,7,\dag}$
}

\date{}
    
\makeatletter
\let\saved@includegraphics\includegraphics
\AtBeginDocument{\let\includegraphics\saved@includegraphics}

\makeatother

\usepackage{newfloat}

\DeclareFloatingEnvironment[
    fileext=loed,
    listname={Extended Data Figures},
    name={Extended Data Fig.},
    placement=htbp
]{extendedfigure}

\usepackage{booktabs}
\usepackage{multirow}
\usepackage{array}
\usepackage{graphicx}
\usepackage{pdflscape}

\begin{document}
%TC:ignore
\maketitle
\begin{affiliations}

\item School of Physical Science and Technology, Beijing University of Posts and Telecommunications, Beijing 100876, China

\item School of Artificial Intelligence, University of Chinese Academy of Sciences, Beijing 100049, China

\item International Institute of Brain-Machine Intelligence, Harbin Institute of Technology, Shenzhen 518055, China

\item School of Biomedical Engineering, Harbin Institute of Technology, Shenzhen 518055, China

\item Beijing Key Laboratory of Brainnetome and Brain-Computer Interface, Institute of Automation, Chinese Academy of Sciences, Beijing 100190, China

\item Brainnetome Center, Institute of Automation, Chinese Academy of Sciences, Beijing 100190, China

\item Xiaoxiang Institute for Brain Health and Yongzhou Central Hospital, Yongzhou, Hunan 425000, China

\item School of Mathematical Sciences, Peking University, Beijing 100871, China

\item Key Laboratory of Mathematics and Information Networks (Beijing University of Posts and Telecommunications), Ministry of Education, Beijing 100876, China

\item Hupan Lab, Hangzhou 310023, China

\item[$^{\ast}$] Contributed equally (Co-first)
\item[$^{\dag}$] Co-senior authors \\
\textbf{Lead Contact}: \\
Tianzi Jiang (jiangtz@nlpr.ia.ac.cn)

\end{affiliations}

% \clearpage
\Heading{Abstract}

\begin{spacing}{1.2}
\noindent

\textbf{
Single-cell-resolution spatial transcriptomics (scST) preserves tissue architecture but often provides targeted or sparse transcriptomic measurements, whereas scRNA-seq offers broader coverage without spatial context.
We present SpCAST, a scalable and interpretable framework that uses scRNA-seq references to transfer cell identity, reconstruct expression and expose gene-level decision evidence in scST.
SpCAST jointly learns reference-cell classification, reference--query alignment and query reconstruction in mini-batches, avoiding the need for a global reference-by-query correspondence matrix.
Spatially Aware Gene Attribution (SAGA) approximates the learned decision function with a sparse additive Kolmogorov--Arnold network.
Across 53 sections comprising 413,404 spatial cells from five technologies, SpCAST achieved the highest aggregate annotation rank among seven methods and scaled to ten million simulated cells.
Controlled masking recovered cell-type-associated expression signals and improved spatial marker concordance.
SAGA further resolved expression-dependent gene evidence and distinguished evidence retained or attenuated across intra- and cross-species reference settings.
}

\end{spacing}

% \newpage
%%%%%%%%%%%%%%%%%%%%%%%%%%%
% Section 1: Introduction %
%%%%%%%%%%%%%%%%%%%%%%%%%%%
% \detailtexcount{main} 
%TC:endignore
%\section{Main}

\clearpage
\begin{spacing}{1.35}
% \linenumbers
\Heading{Introduction}

Single-cell RNA sequencing (scRNA-seq) resolves cellular diversity with near-transcriptome-wide coverage, but tissue dissociation removes cells from their native spatial and microenvironmental context~\cite{Chen2018TissuesCellTypes}. By contrast, single-cell-resolution spatial transcriptomic technologies preserve tissue architecture, enabling individual cells to be profiled in situ~\cite{Stahl2016SpatialTranscriptomics,Codeluppi2018OsmFISH}. Yet spatial resolution does not guarantee molecular completeness: imaging-based platforms are typically restricted to predefined gene panels~\cite{Chen2015MERFISH,Lubeck2014SeqFISH,Eng2019SeqFISHPlus,He2022SpatialMolecularImaging,Janesick2023Xenium}, whereas sequencing-based approaches often remain sparse at cellular resolution~\cite{Russell2024SlideTagsNature,Chen2022StereoSeqOrganogenesis}. Integrating the molecular breadth and detailed cell-type annotations of scRNA-seq references with sparse or targeted spatial measurements therefore offers a route to infer spatial cell identities and recover sparse cell-type-associated expression signals.

A growing number of computational methods integrate single-cell references with spatial transcriptomic data for spatial cell-type annotation, cross-modal alignment and expression reconstruction~\cite{Kleshchevnikov2022Cell2location,Biancalani2021Tangram,Shen2022SpatialID,Wan2023SpatialScope,Hao2024STEM,Yuan2024SPANN,Fan2023SPASCER}. Yet most are optimized around a primary inference objective, leaving cell-type transfer or expression reconstruction to separate models, components or downstream procedures~\cite{Li2022BenchmarkingIntegration}. A separate challenge is computational scale. As spatial profiling expands from restricted fields of view to large tissue sections, whole organs, multi-sample cohorts and cellular atlases~\cite{Yao2023WholeMouseBrainAtlas,Zhang2023WholeMouseBrainMERFISH,Langlieb2023MouseBrainCytoarchitecture}, while reference datasets grow in both size and taxonomic resolution~\cite{TabulaSapiens2022,Siletti2023HumanBrainCellTypes}, methods that construct complete reference-to-query mapping matrices, perform iterative posterior inference or repeatedly aggregate information over cellular neighbourhood graphs can incur substantial computational and memory costs. These computational considerations further motivate a scalable reference-guided framework that combines cell-type transfer and expression-signal recovery without explicitly constructing a complete reference--query mapping.

Accurate and efficient prediction alone, however, does not reveal the molecular evidence underlying an integration model's decisions. Cell-type labels, posterior probabilities, mapping matrices and latent representations describe what a model predicts, but do not directly explain why an individual spatial cell is assigned to a particular identity. Existing interpretable approaches provide complementary views of model behaviour through pathways, regulons, gene programs, individual-gene importance scores, or spatial expression factors and gradients~\cite{chen2023transformer,zhong2024interpretable,chitra2025mapping,gold2026scoring,LundbergLee2017SHAP}. However, these representations do not necessarily provide an explicit decomposition of a specific class decision for an individual spatial cell into expression-dependent gene contributions. Linking a predicted spatial identity to such gene-level decision evidence and localizing that evidence within tissue therefore remains challenging. Such decision-centred and spatially resolved interpretation is particularly relevant in targeted, sparse, cross-platform or cross-species settings, where unexpected predictions may reflect biological variation, incomplete panel coverage, reference--query mismatch, technical effects or model uncertainty.

Here, we present SpCAST, a unified, scalable and interpretable framework for reference-guided integration of scRNA-seq and single-cell-resolution spatial transcriptomic data. Its teacher model jointly learns reference-cell classification, reference--query distribution alignment and spatial-query expression reconstruction, thereby connecting cell-type transfer and expression recovery through a shared latent representation. SpCAST operates on independently sampled mini-batches and avoids explicitly constructing a complete reference-by-query cell-mapping matrix, supporting application to large spatial datasets. To expose the molecular decision evidence associated with the teacher's predictions, we developed Spatially Aware Gene Attribution (SAGA), which uses knowledge distillation~\cite{Hinton2015Distilling,LopezPaz2016Generalized} to train a sparse, single-layer additive Kolmogorov--Arnold network surrogate~\cite{Liu2024KAN} that approximates the teacher's decision function. In SAGA, each class logit is represented as a sum of gene-specific nonlinear response functions. These functions are further summarized as cell-type-specific marker scores and evaluated in individual spatial cells to generate spatial evidence maps.

We evaluated SpCAST across 53 single-cell-resolution spatial transcriptomic sections comprising 413,404 spatial cells from five technologies and spanning multiple tissues and species~\cite{Eng2019SeqFISHPlus,Moffitt2018Hypothalamus,He2022SpatialMolecularImaging,Russell2024SlideTagsNature,Janesick2023Xenium}. Among the seven evaluated methods, SpCAST achieved the highest aggregate cell-type annotation rank, remained stable across random initializations and showed favourable runtime, while scaling to ten million simulated spatial cells. In controlled masking experiments, the same framework recovered cell-type-associated expression signals and increased the spatial concordance of selected marker-gene patterns with their corresponding annotated cell populations. SAGA resolved expression-dependent gene responses and spatial decision evidence associated with cell-type predictions in hypothalamic and cortical sections, including subtype-specific evidence across cortical layers. In a Xenium breast tumour section, SpCAST further provided candidate identities for originally unlabelled cells, which were supported by the expression of cell-type-associated marker genes. When the same human spatial query was analysed using human and ortholog-mapped mouse references, SAGA distinguished gene-response patterns retained across the two reference settings from those attenuated after substitution with the mouse reference. Together, SpCAST establishes a common framework connecting scalable spatial cell-type transfer and expression-signal recovery with spatially resolved gene-level decision evidence.

\Heading{Results}\label{sec3}
\begin{figure*}[!t]
\centering
\includegraphics[width=\textwidth,keepaspectratio]{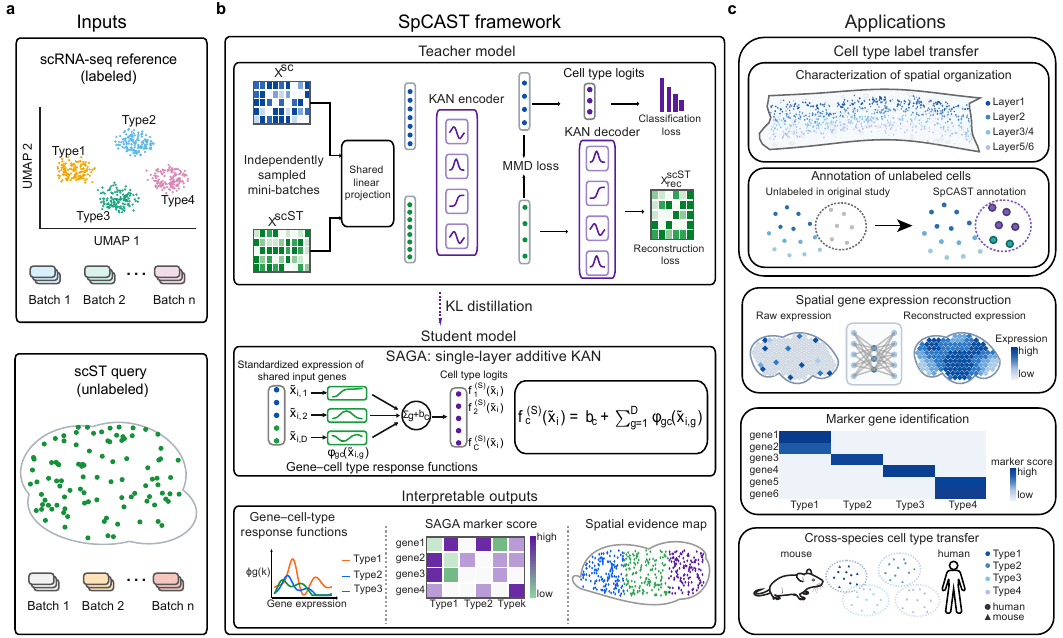}
\caption{\textbf{Overview of the SpCAST framework.}
\textbf{a,} SpCAST takes a labelled scRNA-seq reference and an unlabelled single-cell-resolution spatial transcriptomic query, restricts the two modalities to a common input-gene set and samples reference and query cells independently for joint mini-batch training.
\textbf{b,} A shared linear projection and Kolmogorov--Arnold network (KAN) encoder map reference and query cells into a common latent space. The teacher jointly optimizes supervised reference-cell classification, maximum mean discrepancy (MMD)-based cross-modal distribution alignment and query-specific expression reconstruction. The teacher decision function is subsequently distilled into SAGA, a sparse, single-layer additive KAN surrogate that provides gene--cell-type response functions, SAGA marker scores and spatial decision-evidence maps.
\textbf{c,} SpCAST supports three broad analysis capabilities: spatial cell-type transfer, SAGA-based gene-level interpretation and spatial expression reconstruction. Transferred cell-type predictions enable characterization of spatial cellular organization and candidate assignment of originally unlabelled cells, whereas SAGA enables marker-gene prioritization and comparison of gene-level decision evidence across reference settings, including cross-species transfer.}
\label{fig:fig1}
\end{figure*}

\heading{Overview of the SpCAST framework}
\label{sec:results_overview}

SpCAST integrates a labelled scRNA-seq reference with an unlabelled single-cell-resolution spatial transcriptomic query in a shared gene-expression  space (Fig.~\ref{fig:fig1}a). Reference and query cells are sampled independently  into mini-batches and jointly used for model optimization, without requiring  one-to-one correspondence between cells from the two modalities.

SpCAST comprises a predictive teacher and an interpretable student (Fig.~\ref{fig:fig1}b). In the teacher model, a shared linear projection and KAN encoder~\cite{Liu2024KAN} map reference and spatial query cells into a common latent space. Supervised classification of the reference cells preserves cell-type-discriminative information, maximum mean discrepancy~\cite{Gretton2012MMD} encourages alignment between the latent distributions of the two modalities, and a KAN decoder applied only to the spatial query branch reconstructs expression profiles from the query-cell embeddings. By combining these objectives in mini-batch training, the teacher jointly produces spatial cell-type predictions and reconstructed expression profiles without explicitly constructing a complete reference-by-query cell-mapping matrix.

To expose the gene-level evidence underlying these predictions, SpCAST distils the teacher's decision function~\cite{Hinton2015Distilling} into Spatially Aware Gene Attribution (SAGA), a sparse, single-layer additive KAN model~\cite{Liu2024KAN}. SAGA represents each cell-type logit as a sum of gene-specific nonlinear response functions, revealing how the direction and magnitude of each gene's contribution vary across its expression range. These functions yield cell-type-specific response curves and SAGA marker scores and can be evaluated in individual spatial cells to generate spatial evidence maps. Together, SpCAST links cell-type transfer and expression reconstruction to model-derived marker-gene prioritization, spatial localization of cell-identity decision evidence, cross-species comparison of gene-response functions and marker-supported candidate identities for previously unlabelled cells (Fig.~\ref{fig:fig1}c).

% ============================================================
% fig2
% ============================================================
\begin{figure*}[!t]
    \centering
    \includegraphics[
        width=\textwidth,
        keepaspectratio
    ]{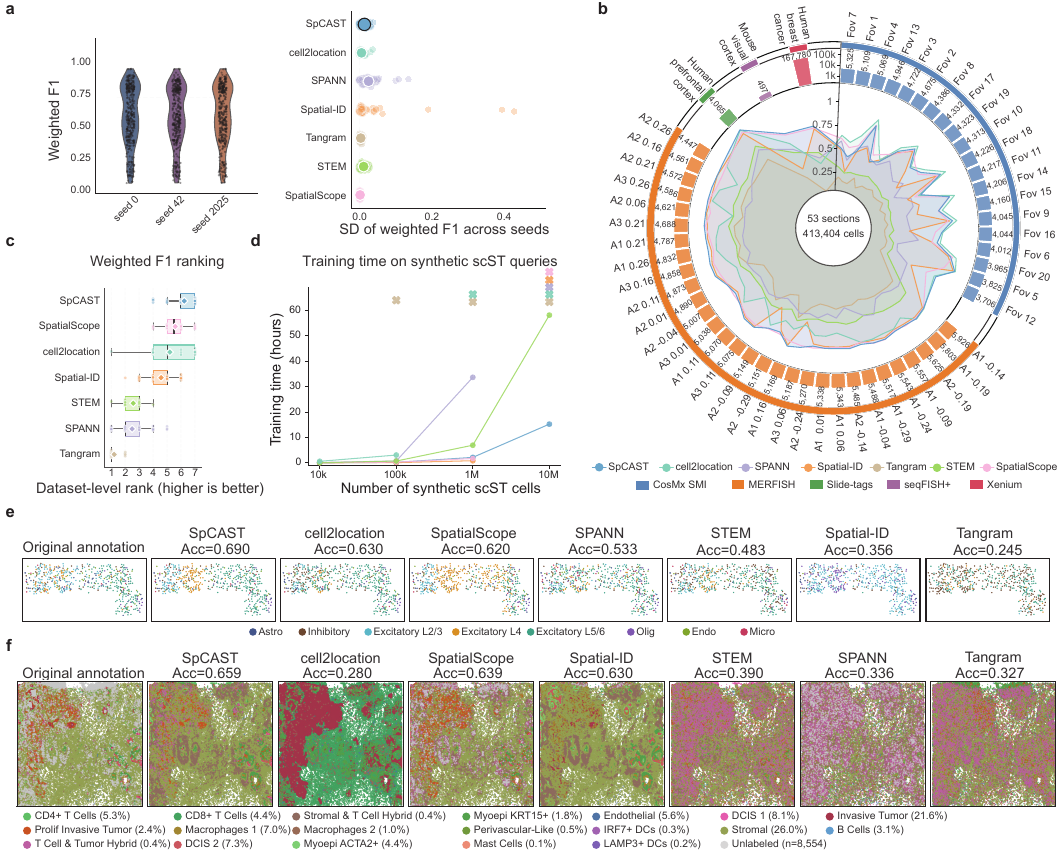}
    \caption{
    \textbf{SpCAST enables accurate, stable and scalable cell-type transfer across single-cell-resolution spatial transcriptomic technologies.}
    \textbf{a,} Stability across random initializations. Left, weighted F1 distributions across 53 spatial sections for three predefined random seeds (0, 42 and 2025). Right, section-level standard deviations of weighted F1 across the three seeds; each point represents one spatial section.
    \textbf{b,} Weighted F1 across 53 benchmark spatial sections. Outer tracks indicate technology, section identity and spatial-cell number, and inner radial profiles show weighted F1 for the seven evaluated methods. The benchmark comprised 413{,}404 spatial cells.
    \textbf{c,} Within-section method ranks based on weighted F1 averaged across the three random seeds; higher ranks indicate better relative performance. Centre lines indicate medians, boxes the interquartile range and whiskers 1.5 times the interquartile range.
    \textbf{d,} Training time on synthetic scST queries ranging from 10{,}000 to 10 million spatial cells. Crosses denote runs that did not complete under the predefined computational-resource constraints.
    \textbf{e,} Original annotations and predicted spatial cell types in the mouse visual cortex seqFISH+ dataset for random seed 2025. Predictions from SpCAST and six comparison methods are shown; values above the maps indicate accuracy for the displayed runs.
    \textbf{f,} Original annotations and predicted spatial cell types in the human breast cancer Xenium dataset for random seed 2025. Predictions from SpCAST and six comparison methods are shown; values above the maps indicate accuracy for the displayed runs. Cells originally designated as Unlabeled were excluded from accuracy calculation.
    }
    \label{fig:benchmark}
\end{figure*}

% ============================================================
% Results part 2 
% ============================================================

\heading{SpCAST enables accurate and scalable cell-type transfer across single-cell-resolution spatial transcriptomic technologies}
\label{subsec_results_benchmark}

To systematically evaluate reference-guided cell-type transfer, we evaluated 53 single-cell-resolution spatial transcriptomic sections comprising 413{,}404 spatial cells across five technologies: CosMx SMI, MERFISH, Slide-tags, seqFISH+ and Xenium~\cite{He2022SpatialMolecularImaging,Moffitt2018Hypothalamus,Russell2024SlideTagsNature,Eng2019SeqFISHPlus,Janesick2023Xenium}(Supplementary Table~\ref{tab:paired_scrna_scst_datasets}). We compared SpCAST with six established methods for scRNA-seq and spatial transcriptomic integration or reference-guided annotation: cell2location~\cite{Kleshchevnikov2022Cell2location}, SPANN~\cite{Yuan2024SPANN}, Spatial-ID~\cite{Shen2022SpatialID}, Tangram~\cite{Biancalani2021Tangram}, STEM~\cite{Hao2024STEM} and SpatialScope~\cite{Wan2023SpatialScope}. Each method was evaluated using the default or recommended configuration of its official implementation across three predefined random seeds, with query annotations reserved for final performance evaluation. Cell-type transfer performance was quantified using accuracy and weighted F1.

Across the benchmark, SpCAST achieved the highest cross-section ranking based on weighted F1 (Fig.~\ref{fig:benchmark}b,c). An independent ranking based on accuracy also placed SpCAST first (Extended Data Fig.~\ref{fig:benchmark_resources}b), indicating that its aggregate performance was not driven by a single evaluation measure. Technology-level accuracy and weighted F1 summaries are provided in Supplementary Table~\ref{tab:real_tech_summary}. A corresponding comparison between the KAN-based SpCAST model and the dimension-matched MLP variant is shown in Supplementary Fig.~\ref{fig:supp_kan_mlp_ablation}. SpCAST predictions were also stable across random initializations: weighted F1 distributions were similar across the three predefined random seeds, and dataset-level standard deviations of weighted F1 across seeds were generally small (Fig.~\ref{fig:benchmark}a).

The aggregate benchmark results were reflected in representative tissues with distinct architectures and cellular compositions. In the mouse visual cortex dataset, SpCAST achieved the highest accuracy among the evaluated methods (0.690), compared with 0.630 for cell2location and 0.620 for SpatialScope, and recapitulated the major spatial patterns in the original annotation, including the laminar organization of excitatory neuronal populations~\cite{Eng2019SeqFISHPlus,Tasic2016MouseV1Taxonomy,Tasic2018NeocorticalAreas,Yao2021IsocortexHPF} (Fig.~\ref{fig:benchmark}e). In the human Xenium breast cancer dataset, SpCAST again achieved the highest accuracy (0.659), exceeding SpatialScope (0.639) and Spatial-ID (0.630), while preserving the broad spatial organization of the major tumour, stromal and immune populations~\cite{Janesick2023Xenium,Wu2021BreastCancerAtlas,Kumar2023HumanBreastAtlas,Reed2024HumanBreastCellAtlas} (Fig.~\ref{fig:benchmark}f).

To compare computational scalability under controlled conditions, we generated synthetic scRNA-seq references and scST queries of increasing size. For methods exposing batch-size and training-epoch parameters, we used a standardized computational budget of 64 cells per mini-batch and 100 training epochs. As the spatial query increased from 10{,}000 cells to 10 million cells, SpCAST completed training at every evaluated scale. At the largest scale, only SpCAST and STEM completed successfully within the predefined GPU-memory and system-RAM limits; SpCAST required approximately 15~h, compared with nearly 58~h for STEM (Fig.~\ref{fig:benchmark}d).

Resource profiling further showed that the peak GPU-memory requirement of SpCAST changed only modestly with increasing query size and remained below the available 80-GB GPU-memory limit at 10 million spatial cells. By contrast, several comparison methods failed at larger scales because their GPU-memory or system-memory requirements exceeded the available resources (Extended Data Fig.~\ref{fig:benchmark_resources}a). On the real datasets, using the default or recommended configuration of each method, SpCAST consistently occupied the lower range of the runtime distributions across the five technologies (Extended Data Fig.~\ref{fig:benchmark_resources}c). Because recommended training configurations differed among methods, these measurements reflect practical computational efficiency under routine usage rather than a fully standardized, hardware-normalized speed comparison.

Together, the cross-dataset rankings, stability across random seeds, representative spatial predictions and controlled scalability analyses show that SpCAST provides accurate, stable and scalable cell-type transfer across diverse single-cell-resolution spatial transcriptomic technologies.

% ============================================================
% Results part 3 
% ============================================================

\heading{SpCAST recovers masked cell-type-associated expression and spatial marker patterns}
\label{subsec_results_reconstruction}

We used controlled expression masking to evaluate whether SpCAST could recover reference-defined cell-type-associated expression signals artificially withheld from spatial queries. This benchmark was designed to assess the recovery of biologically informative cell-type-associated signals under controlled perturbation rather than to reproduce the complete technology-specific dropout process of spatial transcriptomics. For each of 53 reference--query sections, the evaluation set comprised the union of the top 20 differentially expressed genes for each reference cell type. Only originally observed nonzero query entries for these genes were eligible for masking and were randomly withheld at fractions of 0.1, 0.3, 0.5 and 0.7, with five masking seeds per fraction. SpCAST, SpaGE~\cite{SpaGE}, stPlus~\cite{stPlus}, Tangram~\cite{Biancalani2021Tangram}, SpatialScope~\cite{Wan2023SpatialScope} and SpaIM~\cite{SpaIM} received the same masked matrix for each dataset--fraction--seed condition. For each run, one Pearson and one Spearman correlation were calculated across all masked cell--gene entries.

% ============================================================
% Figure 3 
% ============================================================

\begin{figure*}[p]
    \centering
    \includegraphics[width=\textwidth,keepaspectratio]{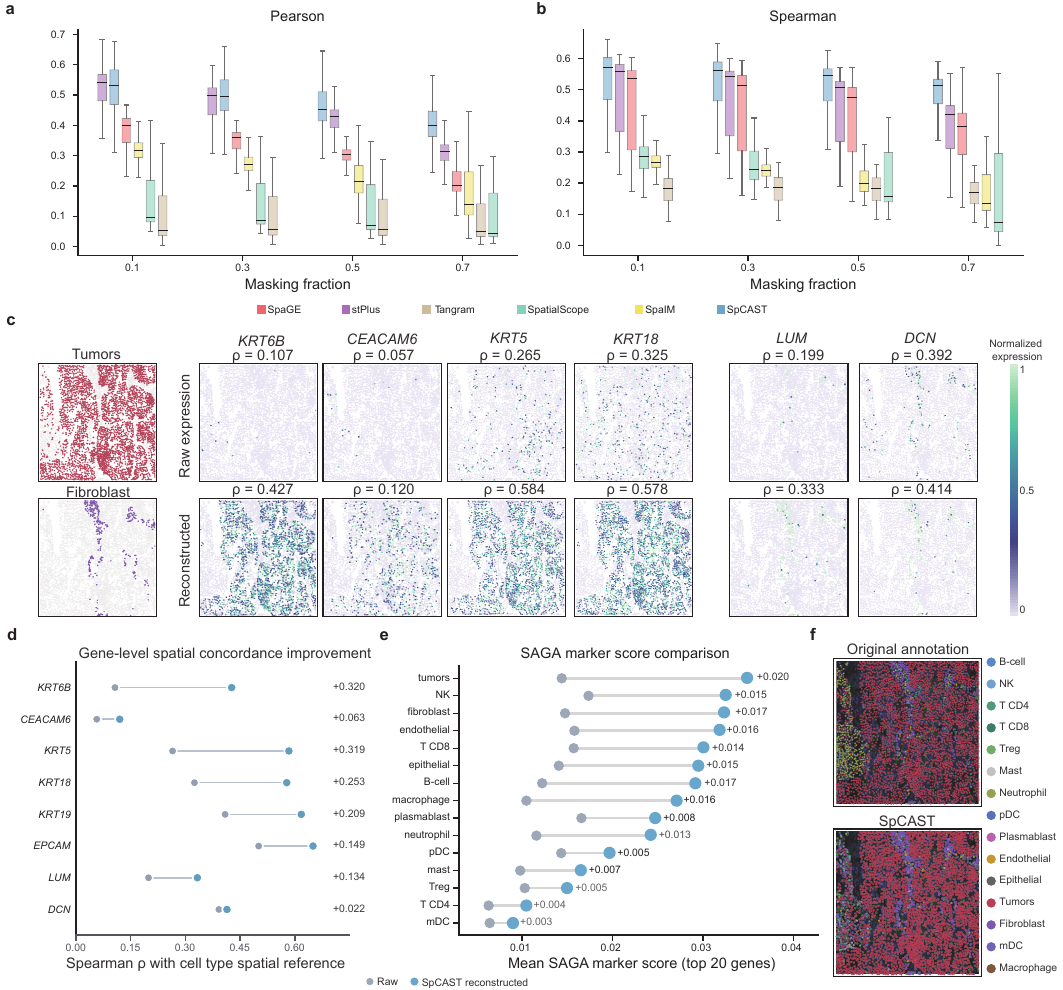}
    \caption{
    \textbf{SpCAST reconstructs masked cell-type-associated expression and increases spatial concordance of marker patterns.}
    \textbf{a,} Pearson correlations between measured and reconstructed values at masking fractions of 0.1, 0.3, 0.5 and 0.7 for six reconstruction methods. Each observation represents one spatial section--masking-seed combination ($n=265$ per method and masking fraction). Centre lines indicate medians, boxes the interquartile range and whiskers 1.5 times the interquartile range.
    \textbf{b,} Spearman correlations for the same comparisons as in \textbf{a}.
    \textbf{c,} Measured and SpCAST-reconstructed expression of six marker genes in representative CosMx SMI FOV 20 (seed 2025), with corresponding tumour-cell and fibroblast spatial references.
    \textbf{d,} Spearman correlations of measured and reconstructed marker expression with the corresponding cell-type spatial references for genes shown in \textbf{c} and Extended Data Fig.~\ref{fig:reconstruction_extended}a. Values indicate reconstructed minus measured correlations.
    \textbf{e,} Mean reference-derived SAGA marker scores across 15 cell populations for models trained independently with measured or reconstructed query expression. The top 20 genes were fixed from the measured-expression setting; values indicate reconstructed-query minus measured-query scores.
    \textbf{f,} Original annotations and SpCAST predictions for CosMx SMI FOV 20 analysed in \textbf{c--e}.
    }
    \label{fig:reconstruction}
\end{figure*}

Reconstruction performance generally declined with increasing masking fraction across methods, but in pooled analyses across the 53 sections, SpCAST remained among the strongest-performing methods at each fraction for both Pearson and Spearman correlation (Fig.~\ref{fig:reconstruction}a,b). Technology-level Pearson and Spearman correlation summaries are provided in Supplementary Table~\ref{tab:marker_tech_summary}. Similar relative patterns were observed across the five masking seeds, indicating that the result was not driven by a single mask realization (Extended Data Fig.~\ref{fig:reconstruction_extended}c). Ablation experiments further assessed the contributions of the scRNA-seq reference and the KAN components. At every masking fraction, the complete SpCAST(KAN) model showed higher median Pearson and Spearman correlations than SpCAST(without sc reference), in which reference supervision and cross-modal alignment were removed, and SpCAST(MLP), in which the KAN components were replaced by multilayer perceptrons (Extended Data Fig.~\ref{fig:reconstruction_extended}b). These results associate both reference information and KAN modelling with stronger reconstruction performance.

We next examined whether reconstruction could strengthen spatial marker signals that were sparse in the measured expression and increase their concordance with the corresponding cell populations. In representative CosMx SMI FOV 20~\cite{He2022SpatialMolecularImaging}, Spearman correlations between four tumour-associated markers and a binary tumour-cell spatial reference derived from the original annotations increased from 0.057--0.325 for measured expression to 0.120--0.584 after reconstruction. Correlations with the fibroblast spatial reference increased from 0.199 to 0.333 for \textit{LUM} and from 0.392 to 0.414 for \textit{DCN}~\cite{Lambrechts2018LungTME,DeZuani2024NSCLCSpatialSingleCell,Luo2022PanCancerCAF} (Fig.~\ref{fig:reconstruction}c,d). Reconstructed marker expression therefore showed greater spatial concordance with the corresponding annotated populations. Within the same CosMx SMI FOV, spatial correlations also increased from 0.409 to 0.618 for \textit{KRT19} and from 0.501 to 0.650 for \textit{EPCAM}~\cite{Lambrechts2018LungTME,DeZuani2024NSCLCSpatialSingleCell} (Extended Data Fig.~\ref{fig:reconstruction_extended}a). As a complementary model-based assessment, we independently trained teacher--SAGA models using either the measured or reconstructed query expression, while retaining the same scRNA-seq reference in both settings. We then compared reference-derived SAGA marker scores between the two models. For each cell type, the top 20 genes ranked by SAGA marker score in the measured-expression setting were fixed across both settings. Mean marker scores for these fixed gene sets were higher in the reconstructed-query setting for all 15 displayed populations, including tumour cells, natural killer cells, fibroblasts and endothelial cells (Fig.~\ref{fig:reconstruction}e,f).

% ============================================================
% Results part 4 
% ============================================================

% \clearpage

\heading{SAGA links spatial cell identities to gene-level decision evidence}
\label{subsec_results_saga_interpretability}

\begin{figure*}[p]
    \centering
    \includegraphics[width=\textwidth,keepaspectratio]{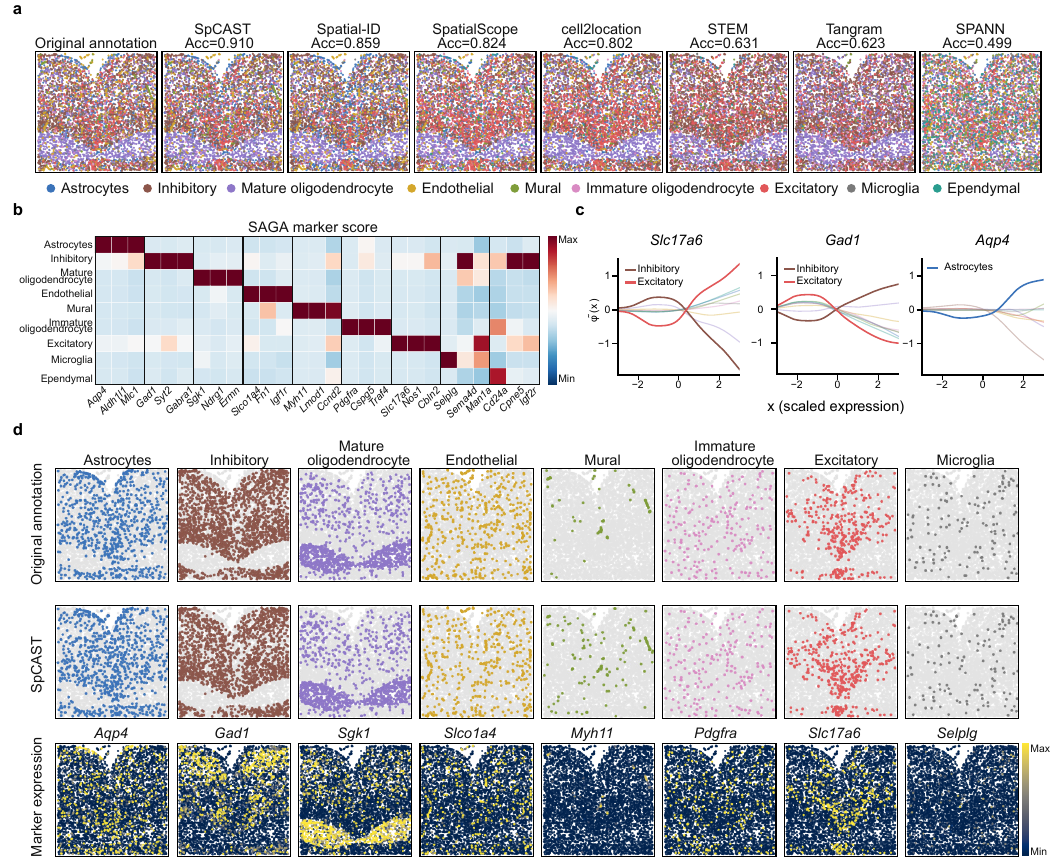}
    \caption{
    \textbf{SAGA links spatial cell-type predictions to gene-level decision evidence in mouse hypothalamus MERFISH data.}
    \textbf{a,} Original cell-type annotations and spatial predictions from SpCAST and six comparison methods for the mouse hypothalamic preoptic-region MERFISH section from animal 1 at Bregma 0.26 mm, using random seed 2025. Values above the maps indicate annotation accuracy for the displayed runs.
    \textbf{b,} SAGA marker-score matrix. Rows denote cell types, columns denote representative high-scoring SAGA genes and colours indicate marker scores for the corresponding gene--cell-type pairs.
    \textbf{c,} Centred SAGA response functions for \textit{Slc17a6}, \textit{Gad1} and \textit{Aqp4}, showing the relationship between standardized expression and gene contributions to the corresponding cell-type decisions. Positive and negative values indicate supportive and suppressive model evidence, respectively.
    \textbf{d,} Original annotations, SpCAST predictions and spatial expression of representative marker genes for eight cell types. The displayed genes are \textit{Aqp4}, \textit{Gad1}, \textit{Sgk1}, \textit{Slco1a4}, \textit{Myh11}, \textit{Pdgfra}, \textit{Slc17a6} and \textit{Selplg}.
    }
    \label{fig:saga_marker}
\end{figure*}

Having established the cell-type transfer and spatial-expression reconstruction capabilities of SpCAST, we next examined the gene-level decision evidence associated with its cell-type predictions. Rather than performing differential-expression analysis after prediction, SAGA approximates the teacher decision function with an additive surrogate whose class logits are explicitly decomposed into gene-specific nonlinear response functions, describing how different expression levels support or suppress individual cell-type decisions. Across the 53 benchmark sections, SAGA showed high top-1 agreement with the SpCAST teacher across spatial transcriptomics technologies (Extended Data Fig.~\ref{fig:saga_fidelity_extended}a). Teacher--student agreement was also stable across the three predefined random seeds (Extended Data Fig.~\ref{fig:saga_fidelity_extended}b), indicating that the additive student consistently reproduced the teacher's top-1 cell-type assignments. We next examined the resulting gene-level decision evidence in a mouse hypothalamic preoptic-region MERFISH dataset containing major neuronal and non-neuronal populations, including astrocytes, inhibitory and excitatory neurons, mature and immature oligodendrocytes, endothelial cells, mural cells, microglia and ependymal cells~\cite{Moffitt2018Hypothalamus,Yao2023WholeMouseBrainAtlas,Zhang2023WholeMouseBrainMERFISH,Campbell2017HypothalamusCensus}.

In the MERFISH section from animal 1 at Bregma 0.26 mm, SpCAST achieved an accuracy of 0.910, exceeding Spatial-ID (0.859), SpatialScope (0.824), cell2location (0.802), STEM (0.631), Tangram (0.623) and SPANN (0.499). The SpCAST prediction map retained the major spatial distributions of neuronal and non-neuronal populations in the original annotation, providing a basis for examining the correspondence between SAGA-derived gene evidence and spatial cell identity (Fig.~\ref{fig:saga_marker}a).

The SAGA marker-score matrix revealed distinct high-scoring gene combinations for different cell types. Astrocyte-associated evidence included \textit{Aqp4}, \textit{Aldh1l1} and \textit{Mlc1}; inhibitory-neuron evidence included \textit{Gad1}, \textit{Syt2} and \textit{Gabra1}; mature-oligodendrocyte evidence included \textit{Sgk1}, \textit{Ndrg1} and \textit{Ermn}; endothelial and mural-cell decisions were associated with \textit{Slco1a4} and \textit{Myh11}, respectively; and immature oligodendrocytes, excitatory neurons, microglia and ependymal cells were associated with \textit{Pdgfra}, \textit{Slc17a6}, \textit{Selplg} and \textit{Cd24a}, respectively (Fig.~\ref{fig:saga_marker}b). Thus, SAGA did not assign a common set of highly expressed genes to all populations, but produced cell-type-specific patterns of model-derived evidence consistent with established neural and glial transcriptional programs~\cite{Cahoy2008GlialTranscriptome,Zeisel2018MouseNervousSystem,Saunders2018DropVizMouseBrain,CellMarker2023,PanglaoDB2019,CellMarker2019}.

The gene-response functions further showed that SAGA provides more than a ranking of candidate markers by resolving the directional relationship between expression and model contribution. Increasing \textit{Slc17a6} expression strengthened positive evidence for excitatory-neuron decisions, whereas its contribution to inhibitory-neuron decisions was weaker or suppressive. Increasing \textit{Gad1} expression preferentially supported inhibitory-neuron decisions, and increasing \textit{Aqp4} expression supported astrocyte decisions (Fig.~\ref{fig:saga_marker}c). Because these response functions are learned directly by SAGA, they indicate not only whether a gene contributes to a cell-type decision, but also the expression range over which supportive or suppressive evidence emerges.

Spatial expression patterns provided further biological support for the model-derived gene-level evidence. Regions of elevated \textit{Aqp4}, \textit{Gad1}, \textit{Sgk1}, \textit{Slco1a4}, \textit{Myh11}, \textit{Pdgfra}, \textit{Slc17a6} and \textit{Selplg} expression were visually concordant with the spatial distributions of their corresponding cell types in both the original annotations and the SpCAST predictions (Fig.~\ref{fig:saga_marker}d). Additional SAGA-prioritized genes showed corresponding spatial expression patterns across these cell populations (Supplementary Fig.~\ref{fig:supp_saga_additional_marker_maps}). These results indicate that high-scoring SAGA genes distinguish cell identities at the model level while retaining spatial expression patterns compatible with the corresponding cellular populations.

We next examined whether this gene-level interpretation generalized to transcriptionally related neuronal subtypes in mouse visual cortex seqFISH+ data. SpCAST retained the relative spatial organization of excitatory L2/3, L4 and L5/6 neurons. \textit{Cux2} and \textit{Ngb} localized predominantly to the L2/3-associated region, \textit{Rorb} and \textit{Scnn1a} to the L4-associated region, and \textit{Fezf2} and \textit{Foxp2} to deeper L5/6-associated regions. The corresponding SAGA evidence maps showed subtype-specific supportive or suppressive evidence across these laminar domains, and the response functions for \textit{Cux2}, \textit{Rorb}, \textit{Fezf2} and \textit{Foxp2} linked marker expression to the respective subtype decisions (Extended Data Fig.~\ref{fig:visual_cortex_extended})~\cite{Eng2019SeqFISHPlus,Tasic2016MouseV1Taxonomy,Tasic2018NeocorticalAreas,Belgard2011NeocorticalLayers,Yao2021IsocortexHPF,Molyneaux2007CorticalSubtype,Nieto2004CuxGenes,Cubelos2010Cux1Cux2,Clark2020RORB,Miskic2021CUX2,Kast2019FOXP2}.

Finally, we evaluated SpCAST candidate identities for originally unlabelled cells in a human breast cancer Xenium dataset using cell-type-associated marker-gene expression. Of 8{,}554 cells originally designated as Unlabeled, 4{,}573 cells with more than 10 total transcripts and more than 5 detected genes passed transcript-signal quality control and were retained for candidate-identity analysis. Representative marker-gene expression patterns across the originally annotated non-Unlabeled cell populations are shown in Supplementary Fig.~\ref{fig:supp_xenium_marker_heatmap}. Among these QC-passed cells, SpCAST candidate assignments showed cell-type-specific marker support: cells assigned to stromal or stromal/T-cell-hybrid categories expressed \textit{LUM}, \textit{POSTN} and \textit{SFRP4}; candidate Myoepithelial ACTA2+ cells expressed \textit{ACTA2}, \textit{MYLK} and \textit{KRT14}; and candidate B cells expressed \textit{MS4A1}, \textit{CD79A} and \textit{CD79B}. These markers were more highly expressed in their respective target candidate populations than in the remaining QC-passed cells, providing further molecular support for the corresponding SpCAST candidate assignments (Extended Data Fig.~\ref{fig:unlabeled_xenium_extended})~\cite{Janesick2023Xenium,Wu2021BreastCancerAtlas,Kumar2023HumanBreastAtlas,Reed2024HumanBreastCellAtlas,Gray2022HumanBreastAtlas,Azizi2018BreastImmuneMap,Jackson2020BreastPathologyLandscape,Nguyen2018BreastEpithelial,BhatNakshatri2021HealthyBreast,Xu2024BreastTumorAtlas,Yang2024PanCancerBCells}. Additional marker-supported candidate populations, including Macrophages 1, endothelial cells, perivascular-like cells and CD8+ T cells, are shown in Supplementary Fig.~\ref{fig:supp_xenium_marker_validation}.

Together, analyses of mouse hypothalamus and visual cortex data show that SAGA transforms SpCAST cell-type predictions into inspectable gene-response functions, cell-type-specific marker scores and spatially concordant gene evidence, enabling fine-grained interpretation of predicted cell identities. In the human breast cancer data, SpCAST additionally proposed candidate identities for originally unlabelled cells that were supported by marker-gene expression.

% ============================================================
% Results part 5 
% ============================================================

\heading{SpCAST preserves cross-species transfer and reveals species-associated gene evidence}
\label{subsec_results_cross_species}

% ============================================================
% Figure 5
% ============================================================

\begin{figure*}[p]
    \centering
    \includegraphics[width=\textwidth,keepaspectratio]{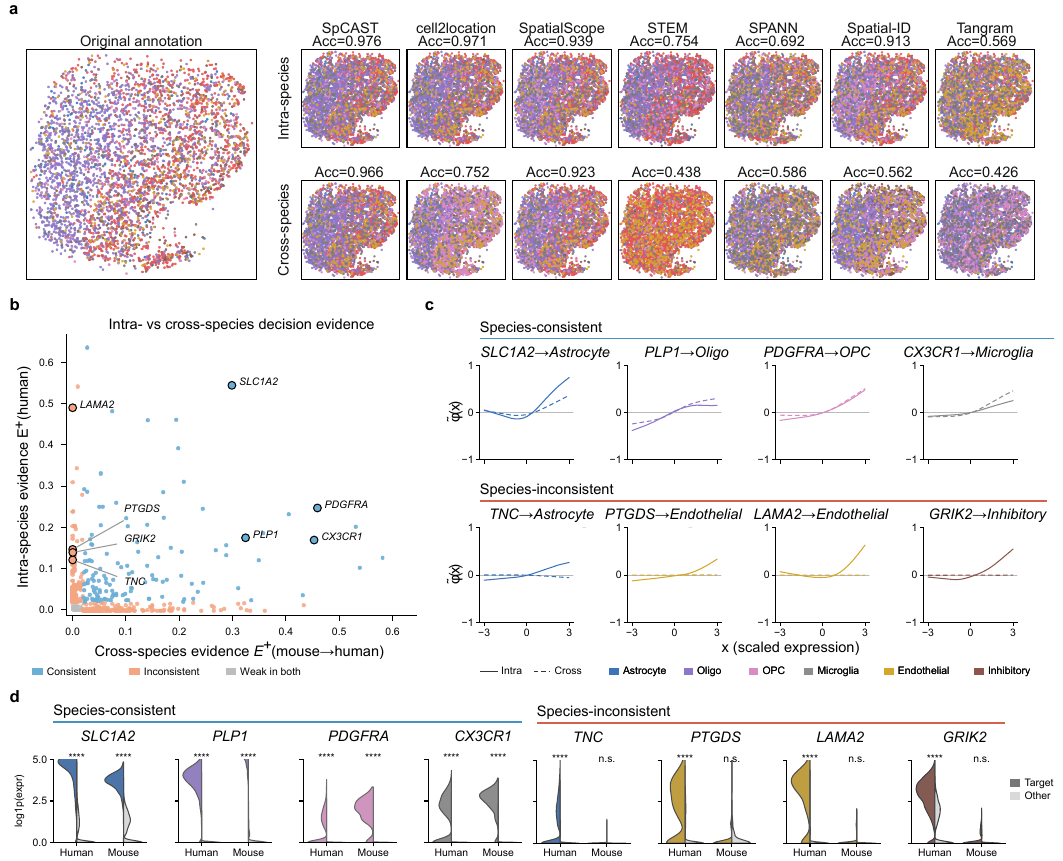}
    \caption{
    \textbf{SpCAST preserves cross-species cell-type transfer while distinguishing species-consistent and species-inconsistent gene evidence.}
    \textbf{a,} Intra-species transfer using a human reference and cross-species transfer to the same human Slide-tags query using an ortholog-mapped mouse reference, shown for random seed 2025. Original annotations and predictions from SpCAST and six comparison methods are shown; values above the maps indicate accuracy for the displayed runs.
    \textbf{b,} Positive decision evidence, \(E^{+}\), for shared gene--cell-type pairs in the intra-species and cross-species models. Each point represents one gene--cell-type pair and is operationally classified as species-consistent, species-inconsistent or weak in both settings according to the evidence-retention criteria defined in the Methods. Selected genes are labelled.
    \textbf{c,} Centered SAGA response functions for representative species-consistent and species-inconsistent genes. Solid lines denote the intra-species model and dashed lines denote the cross-species model.
    \textbf{d,} Expression distributions of representative genes in the human and mouse references. Target denotes the corresponding target cell type and Other denotes all remaining reference cells. Wilcoxon rank-sum tests; $^{****}P<0.0001$; n.s., not significant.
    }
    \label{fig:cross_species}
\end{figure*}

To evaluate reference-guided transfer when the reference and spatial query originated from different species, we constructed intra-species and cross-species analyses using the same human Slide-tags spatial transcriptomic query~\cite{Russell2024SlideTagsNature}. In the intra-species setting, a human prefrontal cortex snRNA-seq dataset was used as the reference~\cite{Herring2022HumanPFC}. In the cross-species setting, genes in a mouse prefrontal cortex scRNA-seq reference were mapped to their corresponding human orthologs before annotation of the same human spatial query~\cite{Bhattacherjee2019MousePFC}. The human Slide-tags query and ortholog-mapped mouse reference shared 13,281 genes, with 15,287 genes unique to the human query and 6,236 unique to the mouse reference (Extended Data Fig.~\ref{fig:cross_species_extended}h).

In intra-species transfer, SpCAST achieved an accuracy of 0.976, comparable to cell2location (0.971) and higher than SpatialScope (0.939), Spatial-ID (0.913), STEM (0.754), SPANN (0.692) and Tangram (0.569). When the mouse reference was used to annotate the same human spatial query, SpCAST retained an accuracy of 0.966, corresponding to an absolute decrease of 0.010 relative to the intra-species setting. By comparison, cross-species accuracies decreased to 0.752 for cell2location, 0.923 for SpatialScope, 0.562 for Spatial-ID, 0.586 for SPANN, 0.438 for STEM and 0.426 for Tangram. The SpCAST cross-species prediction map also visually retained the major cell-type distributions present in the original annotations (Fig.~\ref{fig:cross_species}a).

We next used SAGA to compare gene-level decision evidence between the intra-species and cross-species models. For each shared gene and corresponding cell type, we quantified its positive decision-evidence score, \(E^{+}\), in both settings and operationally classified gene--cell-type pairs as species-consistent, species-inconsistent or weak in both settings according to criteria described in the Methods. Species-consistent genes showed strong cell-type evidence in both reference settings, whereas species-inconsistent genes exhibited strong evidence predominantly in one setting. These labels describe evidence-retention patterns across the two reference settings and do not imply that the observed differences are attributable exclusively to species. \textit{SLC1A2}, \textit{PLP1}, \textit{PDGFRA} and \textit{CX3CR1} showed strong cell-type evidence in both reference settings for astrocytes, oligodendrocytes, OPCs and microglia, respectively. By contrast, \textit{TNC}, \textit{PTGDS}, \textit{LAMA2} and \textit{GRIK2} showed marked differences in decision evidence between the two settings (Fig.~\ref{fig:cross_species}b).

The SAGA response functions further resolved the expression dependence of these differences. For the species-consistent genes, \textit{SLC1A2}, \textit{PLP1}, \textit{PDGFRA} and \textit{CX3CR1} exhibited directionally similar cell-type responses in the intra- and cross-species models, with increasing expression associated with stronger positive contributions to the corresponding cell-type decisions. In contrast, \textit{TNC}, \textit{PTGDS}, \textit{LAMA2} and \textit{GRIK2} showed pronounced positive responses in the intra-species model but reduced response amplitudes or comparatively flat functions in the cross-species model (Fig.~\ref{fig:cross_species}c). SAGA therefore distinguished gene-response patterns that were retained across the two reference settings from decision evidence that weakened under a heterologous reference.

Expression distributions in the human and mouse references provided further support for this classification. Species-consistent genes showed significant target-cell-type enrichment in both reference datasets: \textit{SLC1A2} in astrocytes, \textit{PLP1} in oligodendrocytes, \textit{PDGFRA} in OPCs and \textit{CX3CR1} in microglia were more highly expressed in their target populations than in other cell types in both the human and mouse references. By contrast, \textit{TNC}, \textit{PTGDS}, \textit{LAMA2} and \textit{GRIK2} showed significant target-cell-type enrichment in the human data but no significant separation in the corresponding mouse populations (Fig.~\ref{fig:cross_species}d). Thus, differences in the learned response functions were consistent with species-associated differences in cell-type-specific expression within the reference datasets.

This pattern extended beyond the genes displayed in the main figure. Additional species-consistent candidates included \textit{ATP1A2} and \textit{CLU} in astrocytes, \textit{MBP} and \textit{CLDN11} in oligodendrocytes, \textit{VCAN} and \textit{NXPH1} in OPCs, \textit{FN1} and \textit{EBF1} in endothelial cells, \textit{CSF1R} and \textit{APBB1IP} in microglia, \textit{ARPP21} and \textit{MEF2C} in excitatory neurons, and \textit{GAD1} and \textit{NRXN3} in inhibitory neurons. Corresponding species-inconsistent candidates generally showed weakened or nonsignificant target-cell-type expression separation in the mouse reference (Extended Data Fig.~\ref{fig:cross_species_extended}a--g). These observations are consistent with previous evidence that homologous brain cell types retain core transcriptional programs while also exhibiting species-associated differences in individual marker genes~\cite{hodge2019conserved,Bakken2021ComparativeMotorCortex,Fang2022MERFISHHumanMouse,Jorstad2023ComparativeTranscriptomics,Krienen2020PrimateInterneurons,Romero2012GeneExpressionEvolution}.

\Heading{Discussion}

In this study, we developed SpCAST, an efficient, scalable and interpretable reference-guided framework for integrating scRNA-seq references with single-cell-resolution spatial transcriptomic data. SpCAST connects spatial cell-type transfer, cross-modal distribution alignment and spatial-expression reconstruction within a common predictive framework, while SAGA provides gene--cell-type-specific nonlinear response functions and spatial decision evidence associated with the resulting cell-type predictions. Together, these components link predicted spatial identities, recovery of cell-type-associated expression signals and gene-level interpretation within a coordinated analytical workflow.

The multi-platform benchmark demonstrates that SpCAST combines strong annotation performance with stable and scalable computation across diverse scST settings. Across 53 sections comprising 413{,}404 spatial cells from five spatial transcriptomics technologies, SpCAST achieved the highest aggregate annotation rank among the seven evaluated methods and remained stable across random initializations. In controlled scaling experiments, SpCAST completed synthetic spatial queries containing up to ten million cells under the tested hardware configuration. This capability is increasingly relevant as spatial studies expand towards large tissue sections, whole organs and atlas-scale cellular maps~\cite{Yao2023WholeMouseBrainAtlas,Zhang2023WholeMouseBrainMERFISH,Langlieb2023MouseBrainCytoarchitecture}. The observed scaling behaviour is consistent with the computational design of the SpCAST teacher, whose classification, alignment and reconstruction objectives are optimized on independently sampled reference and query mini-batches without explicitly constructing a complete reference-by-query cell-mapping matrix. Thus, the principal tensors required for each optimization step are determined by the mini-batch size rather than by a full reference--query correspondence, consistent with the modest increase in peak GPU memory observed as spatial query size increased. These results support mini-batch reference-guided integration as a practical strategy for increasingly large spatial datasets.

Spatial-expression reconstruction represents a second capability that distinguishes SpCAST from conventional label-transfer workflows~\cite{Li2022BenchmarkingIntegration}. Controlled masking of observed expression entries among reference-defined cell-type-associated genes was used to test whether biologically informative signals could be recovered after artificial withholding, rather than to emulate the complete dropout process of a particular spatial transcriptomics technology. Across 53 sections, SpCAST showed strong pooled recovery of these withheld signals across masking levels and random seeds, although reconstruction performance varied across technologies. Comparisons with a query-only variant without the scRNA-seq reference and an MLP-based SpCAST variant further associated reference guidance and KAN-based nonlinear modelling~\cite{Liu2024KAN} with stronger reconstruction performance. In CosMx SMI data~\cite{He2022SpatialMolecularImaging}, reconstructed expression of tumour, epithelial and fibroblast markers showed greater concordance with the corresponding spatial cell-type distributions. In a complementary model-based analysis, models trained with reconstructed query expression showed higher SAGA marker scores across fixed marker-gene sets than models trained with measured query expression. Together, these results indicate that reference-guided reconstruction can recover cell-identity-associated expression signals under controlled masking and strengthen spatial marker patterns associated with the underlying cellular organization.

SAGA provides the gene-level interpretability layer of SpCAST. By approximating the teacher decision function with an additive model, SAGA represents each cell-type score through gene-specific nonlinear responses, allowing the contribution of individual genes to be examined across their expression ranges. Centred response curves distinguish supportive and suppressive gene evidence, whereas marker-score matrices and spatial evidence maps organize this information at gene and tissue levels. This formulation extends the shape-function perspective of generalized additive and neural additive models~\cite{HastieTibshirani1987GAM,Agarwal2021NAM} to spatially resolved cell-type decisions, making the molecular evidence underlying individual predictions directly inspectable across both expression and tissue space. Across brain datasets, SAGA evidence corresponded to established neuronal, glial and cortical laminar marker programmes, connecting predictions of L2/3, L4 and L5/6 excitatory neurons to cortical organization~\cite{Eng2019SeqFISHPlus,Tasic2016MouseV1Taxonomy,Tasic2018NeocorticalAreas,Belgard2011NeocorticalLayers,Yao2021IsocortexHPF,Cahoy2008GlialTranscriptome,Zeisel2018MouseNervousSystem,Molyneaux2007CorticalSubtype,Nieto2004CuxGenes,Clark2020RORB,Kast2019FOXP2}. In Xenium tumour data, SpCAST proposed candidate identities for originally unlabelled cells that passed transcript-signal filtering, with corresponding assignments supported by cell-type-associated marker expression~\cite{Janesick2023Xenium,Wu2021BreastCancerAtlas,Kumar2023HumanBreastAtlas,Reed2024HumanBreastCellAtlas,Gray2022HumanBreastAtlas,Nguyen2018BreastEpithelial,Yang2024PanCancerBCells}. Comparison of response functions learned from human and mouse references further distinguished gene-level decision evidence that was retained or attenuated under cross-species reference transfer. These patterns were concordant with differences in cell-type-specific expression between the reference datasets and with previously reported conserved and divergent transcriptional programmes across human and mouse cortex~\cite{hodge2019conserved,Bakken2021ComparativeMotorCortex,Fang2022MERFISHHumanMouse,Jorstad2023ComparativeTranscriptomics,Krienen2020PrimateInterneurons,Romero2012GeneExpressionEvolution}.

Several limitations remain. First, SpCAST relies on an annotated scRNA-seq or single-nucleus RNA-seq reference, and its transfer performance therefore depends on reference quality, cell-type coverage and annotation resolution~\cite{Li2022BenchmarkingIntegration}. Cell states absent from the reference may consequently be difficult to resolve. Second, targeted gene panels, low detection efficiency and dropout limit the information available for low-transcript cells, rare populations and closely related subtypes~\cite{Janesick2023Xenium,He2022SpatialMolecularImaging,Russell2024SlideTagsNature}. Third, SAGA is an additive surrogate of the teacher, and its response functions represent model-derived decision evidence rather than causal regulation. Extending SpCAST towards open-set annotation, additional molecular or morphological modalities and broader cross-species settings may further expand its applicability.

Together, SpCAST links scalable reference-guided spatial cell-type transfer with recovery of cell-type-associated expression signals and inspectable gene-level decision evidence. By connecting predicted spatial identities, reconstructed cell-type-associated expression and expression-dependent molecular evidence within a common framework, SpCAST provides an interpretable approach for the integrative analysis of sparse and targeted single-cell-resolution spatial transcriptomic data.

\Heading{Methods}
% \section{Methods}\label{sec2}

\heading{Overview of SpCAST}\label{subsec_main_study_design}

SpCAST is a reference-guided teacher--student framework for the joint analysis of a labelled dissociated scRNA-seq or snRNA-seq reference and an unlabelled single-cell-resolution spatial transcriptomic (scST) query. The framework separates predictive modelling from gene-level interpretation and supports spatial cell-type transfer, query-expression reconstruction and gene-level decision-evidence analysis without requiring one-to-one correspondence between reference and query cells.

The SpCAST teacher maps reference and query expression profiles into a shared latent space and jointly learns three objectives. Reference labels provide supervision for cell-type classification, cross-modal distribution alignment reduces discrepancies between the reference and query latent representations, and a query-specific decoder reconstructs spatial expression from the latent representation. After training, the teacher produces cell-type posterior probabilities, predicted labels and reconstructed expression profiles for the query cells.

A subsequent Spatially Aware Gene Attribution (SAGA) model is trained as a strictly additive gene-level surrogate of the pretrained teacher. Each gene--cell-type pair is represented by an explicit univariate response function, allowing the supportive or suppressive contribution of an individual gene to the corresponding surrogate cell-type decision to be quantified. The trained SAGA model is further used to derive gene-to-cell-type response curves, SAGA marker scores and spatial decision-evidence maps.

\heading{SpCAST teacher model}\label{subsec_main_teacher}

\noindent\textbf{Shared representation and model outputs.}

The SpCAST teacher operates on a labelled single-cell reference and an unlabelled scST query represented in a common input-gene space. Let
$\mathbf{X}^{\mathrm{sc}}\in\mathbb{R}^{N_{\mathrm{sc}}\times D}$
and
$\mathbf{y}^{\mathrm{sc}}\in\{1,\ldots,C\}^{N_{\mathrm{sc}}}$
denote the reference expression matrix and its cell-type labels, respectively, and let
$\mathbf{X}^{\mathrm{st}}\in\mathbb{R}^{N_{\mathrm{st}}\times D}$
denote the query expression matrix. Here, $D$ is the number of genes in the shared input panel and $C$ is the number of cell types in the harmonized label vocabulary. Both modalities are represented by log-normalized expression values. For notational convenience, the expression profile of cell $i$ from modality $m$ is treated as a column vector,
$\mathbf{x}^{m}_i=(\mathbf{X}^{m}_{i,:})^\top\in\mathbb{R}^{D}$,
where
$m\in\{\mathrm{sc},\mathrm{st}\}$.

A shared bias-free linear projection first maps each $D$-dimensional expression profile to a 30-dimensional intermediate representation, which is subsequently transformed by a single-layer Kolmogorov--Arnold network (KAN) encoder into a 20-dimensional latent representation:
\begin{equation}
\label{eq:teacher_encoder}
\begin{aligned}
\mathbf{h}^{m}_i
&=
\mathbf{W}_{\mathrm{in}}\mathbf{x}^{m}_i,
\\
\mathbf{z}^{m}_i
&=
\mathcal{E}_{\mathrm{KAN}}
\left(
\mathbf{h}^{m}_i
\right),
\end{aligned}
\qquad
m\in\{\mathrm{sc},\mathrm{st}\}.
\end{equation}
Here,
$\mathbf{W}_{\mathrm{in}}\in\mathbb{R}^{30\times D}$
is the shared input-projection matrix,
$\mathbf{h}^{m}_i\in\mathbb{R}^{30}$
is the intermediate representation and
$\mathbf{z}^{m}_i\in\mathbb{R}^{20}$
is the latent representation of cell $i$.
The encoder therefore defines the mapping
$\mathcal{E}_{\mathrm{KAN}}:\mathbb{R}^{30}\rightarrow\mathbb{R}^{20}$.
The single-cell reference and scST query modalities share the same input-projection and encoder parameters.

The latent representation supports both cell-type classification and query-expression reconstruction. A shared linear classification head produces posterior probabilities over the $C$ cell types for both modalities, whereas a single-layer KAN decoder maps only the query representations back to the input-gene space:
\begin{equation}
\label{eq:teacher_outputs}
\begin{aligned}
\hat{\mathbf{p}}^{m}_i
&=
\operatorname{softmax}
\left(
\mathbf{W}_{\mathrm{cls}}\mathbf{z}^{m}_i
+
\mathbf{b}_{\mathrm{cls}}
\right),
\qquad
m\in\{\mathrm{sc},\mathrm{st}\},
\\
\hat{y}^{\mathrm{st}}_i
&=
\operatorname*{arg\,max}_{c\in\{1,\ldots,C\}}
\hat{p}^{\mathrm{st}}_{ic},
\\
\hat{\mathbf{x}}^{\mathrm{st}}_i
&=
\mathcal{D}_{\mathrm{KAN}}
\left(
\mathbf{z}^{\mathrm{st}}_i
\right).
\end{aligned}
\end{equation}
Here,
$\mathbf{W}_{\mathrm{cls}}\in\mathbb{R}^{C\times 20}$
and
$\mathbf{b}_{\mathrm{cls}}\in\mathbb{R}^{C}$
are the weight matrix and bias vector of the classification head, respectively.
The vector
$\hat{\mathbf{p}}^{m}_i\in[0,1]^C$
contains the predicted cell-type probabilities and satisfies
$\sum_{c=1}^{C}\hat{p}^{m}_{ic}=1$.
The predicted label
$\hat{y}^{\mathrm{st}}_i$
is assigned to the class with the largest posterior probability.
The decoder
$\mathcal{D}_{\mathrm{KAN}}:\mathbb{R}^{20}\rightarrow\mathbb{R}^{D}$
produces the reconstructed query-expression profile
$\hat{\mathbf{x}}^{\mathrm{st}}_i\in\mathbb{R}^{D}$.
Reference representations are not passed through the decoder.

\noindent\textbf{KAN parameterization.}

Both the encoder and decoder are implemented using a single KANLayer\cite{Liu2024KAN}. For an input vector
$\mathbf{u}\in\mathbb{R}^{d_{\mathrm{in}}}$,
the $j$th output of a KAN layer and its corresponding edge functions are defined as
\begin{equation}
\label{eq:kan_layer}
\begin{aligned}
v_j
&=
\sum_{r=1}^{d_{\mathrm{in}}}
\phi_{jr}(u_r),
\qquad
j=1,\ldots,d_{\mathrm{out}},
\\
\phi_{jr}(u)
&=
s^{\mathrm{base}}_{jr}
\operatorname{SiLU}(u)
+
s^{\mathrm{spline}}_{jr}
\sum_{q}
c_{jrq}B_{q,k}(u).
\end{aligned}
\end{equation}
Here,
$\phi_{jr}:\mathbb{R}\rightarrow\mathbb{R}$
is the learnable univariate edge function connecting input dimension $r$ to output dimension $j$,
$s^{\mathrm{base}}_{jr}$
and
$s^{\mathrm{spline}}_{jr}$
are trainable scale parameters,
$c_{jrq}$
is the coefficient of the $q$th B-spline basis function and
$B_{q,k}$
denotes a B-spline basis function of order $k$.

The KAN layers in both the encoder and decoder use five spline-grid intervals, cubic B-splines, a SiLU base function and an initial spline-grid range of $[-1,1]$.

\heading{SpCAST teacher training objectives}
\label{subsec_main_teacher_objectives}

The SpCAST teacher is trained using three complementary objectives. Supervised classification on the reference labels preserves cell-type-discriminative information, maximum mean discrepancy (MMD) aligns the latent distributions of the reference and query, and query-only reconstruction preserves expression information from the spatial modality. Let
$\mathcal{B}_{\mathrm{sc}}$
and
$\mathcal{B}_{\mathrm{st}}$
denote the reference and query mini-batch index sets, respectively, with
$B_{\mathrm{sc}}=|\mathcal{B}_{\mathrm{sc}}|$
and
$B_{\mathrm{st}}=|\mathcal{B}_{\mathrm{st}}|$.
The three component losses are defined as
\begin{equation}
\label{eq:teacher_component_losses}
\begin{aligned}
\mathcal{L}_{\mathrm{CE}}^{\mathrm{sc}}
&=
-\frac{1}{B_{\mathrm{sc}}}
\sum_{i\in\mathcal{B}_{\mathrm{sc}}}
\log
\hat{p}^{\mathrm{sc}}_{i,y^{\mathrm{sc}}_i},
\\[4pt]
\mathcal{L}_{\mathrm{MMD}}^{\mathrm{sc,st}}
&=
\left\|
\frac{1}{B_{\mathrm{sc}}}
\sum_{i\in\mathcal{B}_{\mathrm{sc}}}
\psi\!\left(\mathbf{z}^{\mathrm{sc}}_i\right)
-
\frac{1}{B_{\mathrm{st}}}
\sum_{j\in\mathcal{B}_{\mathrm{st}}}
\psi\!\left(\mathbf{z}^{\mathrm{st}}_j\right)
\right\|_{\mathcal{H}}^{2},
\\[4pt]
\mathcal{L}_{\mathrm{rec}}^{\mathrm{st}}
&=
\frac{1}{B_{\mathrm{st}}D}
\left\|
\hat{\mathbf{X}}_{\mathcal{B}_{\mathrm{st}}}^{\mathrm{st}}
-
\mathbf{X}_{\mathcal{B}_{\mathrm{st}}}^{\mathrm{st}}
\right\|_{\mathrm{F}}^{2}.
\end{aligned}
\end{equation}

\noindent\textbf{Reference-supervised classification.}

Cell-type classification is supervised exclusively by the labelled reference cells. In equation~\eqref{eq:teacher_component_losses},
$\hat{p}^{\mathrm{sc}}_{i,y^{\mathrm{sc}}_i}$
is the probability assigned to the reference label
$y^{\mathrm{sc}}_i$.
The reference classification loss is the mean negative log-likelihood of the correct cell-type labels within the reference mini-batch. Query benchmark labels are not used in this objective.

\noindent\textbf{Latent distribution alignment.}

The MMD term in equation~\eqref{eq:teacher_component_losses} compares the empirical reference and query distributions in the shared 20-dimensional latent space. Here,
$\psi(\cdot)$
is the feature map associated with a reproducing kernel Hilbert space
$\mathcal{H}$.
This objective aligns the marginal latent distributions of the two modalities without requiring one-to-one correspondence between reference and query cells.

The empirical MMD is calculated using five Gaussian kernels with bandwidth set
$\Sigma=\{0.5,1,2,5,10\}$.
For each bandwidth, mean pairwise kernel values are calculated within the reference, between the reference and query and within the query.

\noindent\textbf{Query expression reconstruction.}

The reconstruction term is evaluated only for query cells. In equation~\eqref{eq:teacher_component_losses},
$\mathbf{X}_{\mathcal{B}_{\mathrm{st}}}^{\mathrm{st}}
\in\mathbb{R}^{B_{\mathrm{st}}\times D}$
and
$\hat{\mathbf{X}}_{\mathcal{B}_{\mathrm{st}}}^{\mathrm{st}}
\in\mathbb{R}^{B_{\mathrm{st}}\times D}$
denote the input and reconstructed expression matrices for the current query mini-batch, respectively. Both matrices are represented on the same log-normalized input-gene scale. Division by
$B_{\mathrm{st}}D$
makes
$\mathcal{L}_{\mathrm{rec}}^{\mathrm{st}}$
the mean squared reconstruction error across query cells and input genes. Reference expression profiles are not included in the reconstruction loss.

\noindent\textbf{Joint training objective.}

The three component losses are optimized jointly:
\begin{equation}
\label{eq:teacher_loss}
\mathcal{L}_{\mathrm{teacher}}
=
\lambda_{\mathrm{CE}}
\mathcal{L}_{\mathrm{CE}}^{\mathrm{sc}}
+
\lambda_{\mathrm{MMD}}
\mathcal{L}_{\mathrm{MMD}}^{\mathrm{sc,st}}
+
\lambda_{\mathrm{rec}}
\mathcal{L}_{\mathrm{rec}}^{\mathrm{st}},
\end{equation}
where
$\lambda_{\mathrm{CE}}$,
$\lambda_{\mathrm{MMD}}$
and
$\lambda_{\mathrm{rec}}$
are the loss weights for reference-supervised classification, latent distribution alignment and query expression reconstruction, respectively, with default values of
$1$,
$0.03$
and
$0.01$.

\heading{SAGA distillation and gene-level decision evidence}
\label{subsec_main_saga}

\noindent\textbf{Additive gene-level decision model.}

SAGA approximates the nonlinear decision function of the trained teacher with a strictly additive gene-level surrogate model. Unlike post hoc attribution methods that infer feature importance after prediction, SAGA explicitly parameterizes gene contributions to cell-type decisions within the model architecture.

For a standardized expression profile
$\tilde{\mathbf{x}}_i$,
the SAGA logit for cell type $c$ is defined as
\begin{equation}
\label{eq:saga_additive}
f^{(S)}_c(\tilde{\mathbf{x}}_i)
=
b_c
+
\sum_{g=1}^{D}
\phi_{gc}
\left(
\tilde{x}_{ig}
\right),
\qquad
c=1,\ldots,C,
\end{equation}
where
$b_c$
is the class-specific intercept and
$\phi_{gc}:\mathbb{R}\rightarrow\mathbb{R}$
is the learnable gene--cell-type response function describing the contribution of gene $g$ to the decision of cell type $c$.

Each response function therefore provides an explicit gene-level decomposition of the SAGA surrogate logit.

\noindent\textbf{Teacher--student distillation and regularization.}

SAGA is trained as an additive surrogate of the pretrained teacher rather than as the primary annotation model. Its objective combines teacher--student distillation on reference and spatial query cells with auxiliary reference-label supervision and response-function regularization.

The SAGA objective is
\begin{equation}
\label{eq:saga_objective}
\begin{aligned}
\mathcal{L}_{\mathrm{SAGA}}
={}&
\mathcal{L}_{\mathrm{distill}}^{\mathrm{sc}}
+
\mathcal{L}_{\mathrm{distill}}^{\mathrm{st}}
+
\lambda_{\mathrm{CE}}
\mathcal{L}_{\mathrm{CE}}^{\mathrm{sc}}
\\
&+
\lambda_{L1}
\mathcal{L}_{L1}
+
\lambda_{\mathrm{sparse}}
\mathcal{L}_{\mathrm{group}}
+
\lambda_{\mathrm{smooth}}
\mathcal{L}_{\mathrm{smooth}} .
\end{aligned}
\end{equation}

where
$\mathcal{L}_{\mathrm{distill}}^{\mathrm{sc}}$
and
$\mathcal{L}_{\mathrm{distill}}^{\mathrm{st}}$
denote teacher--student KL divergence on reference and spatial query cells, respectively.
$\mathcal{L}_{\mathrm{CE}}^{\mathrm{sc}}$
is the auxiliary reference-label classification loss.
The
$\mathcal{L}_{L1}$
term applies element-wise edge sparsity,
whereas
$\mathcal{L}_{\mathrm{group}}$
is a group-lasso edge penalty that promotes sparse gene--cell-type response functions.
$\mathcal{L}_{\mathrm{smooth}}$
regularizes spline response smoothness.

The default weights are
$\lambda_{\mathrm{CE}}=0.5$,
$\lambda_{L1}=10^{-3}$,
$\lambda_{\mathrm{sparse}}=10^{-3}$,
and
$\lambda_{\mathrm{smooth}}=10^{-4}$.
The exact definitions of the regularization terms are provided in the Supplementary Methods.

\noindent\textbf{Centred response functions and gene-level evidence.}

To place gene-level decision contributions on a common reference baseline, the learned gene--cell-type response functions were centred after SAGA training. For gene \(g\) and cell type \(c\), the mean response across reference cells was defined as

\begin{equation}
\label{eq:saga_response_mean}
\mu_{gc}
=
\frac{1}{N_{\mathrm{sc}}}
\sum_{i=1}^{N_{\mathrm{sc}}}
\phi_{gc}
\left(
\tilde{x}^{\mathrm{sc}}_{ig}
\right).
\end{equation}

The response function and class-specific intercept were then reparameterized as

\begin{equation}
\label{eq:saga_centered_response}
\widetilde{\phi}_{gc}(t)
=
\phi_{gc}(t)-\mu_{gc},
\qquad
\widetilde{b}_{c}
=
b_{c}
+
\sum_{g=1}^{D}\mu_{gc}.
\end{equation}

This reparameterization preserves the SAGA class-specific logit,

\begin{equation}
\label{eq:saga_centering_invariance}
b_c+
\sum_{g=1}^{D}
\phi_{gc}
\left(
\tilde{x}_{ig}
\right)
=
\widetilde{b}_c+
\sum_{g=1}^{D}
\widetilde{\phi}_{gc}
\left(
\tilde{x}_{ig}
\right).
\end{equation}

Centering therefore changes only the reference baseline of the gene-level contributions, without altering the SAGA decision function. The centred contributions were used to derive gene-level evidence, marker scores and spatial decision-evidence maps.

\noindent\textbf{Spatial decision-evidence mapping.}

For each cell type $c$, a cell-type-specific gene set
$\mathcal{G}^{(q)}_c$
is selected by ranking genes according to their reference-domain SAGA marker scores, where
$q$
is the number of selected genes and defaults to 20.
The spatial decision evidence for query cell $i$ and cell type $c$ is defined as
\begin{equation}
\label{eq:saga_evidence_map}
M_c(i)
=
\sum_{g\in\mathcal{G}^{(q)}_c}
\widetilde{\phi}_{gc}
\left(
\tilde{x}^{\mathrm{st}}_{ig}
\right).
\end{equation}

Positive values of
$M_c(i)$
indicate net supportive evidence for cell type $c$, whereas negative values indicate suppressive evidence.
This formulation maps gene-level decision evidence represented by the SAGA surrogate onto individual spatial cells.

\heading{Model optimization and implementation}
\label{subsec_main_optimization}

\noindent\textbf{Teacher optimization.}

During each optimization step, a query mini-batch was first sampled from the scST query with a default batch size of 256. The same number of cells was then sampled with replacement from the single-cell reference, ensuring matched reference and query mini-batch sizes. Reference cells were sampled using inverse-log frequency weighting to reduce the dominance of highly abundant cell types during optimization.

Specifically, cells from reference cell type $c$ were assigned sampling weights proportional to

\begin{equation}
w_c
=
\frac{1}{\log(1+n_c)},
\end{equation}

where
$n_c$
denotes the number of reference cells belonging to cell type $c$.
The resulting sampling probability reduces class-frequency imbalance without enforcing exact class-balanced sampling.

By default, the teacher was optimized using Adam for 10 epochs with a learning rate of $10^{-3}$, weight decay of $10^{-4}$, and global gradient clipping at 5. For the seqFISH+ dataset, which comprised only 497 spatial cells, training was extended to 30 epochs to increase the number of optimization steps.

\noindent\textbf{SAGA optimization.}

Following teacher training, SAGA was optimized with the teacher parameters held fixed. Teacher soft probabilities on reference and spatial query cells served as distillation targets, while reference labels provided auxiliary supervision through the classification loss. SAGA was therefore optimized as an interpretable additive surrogate of the pretrained teacher rather than as the primary annotation model.

Because SAGA learns gene-level decision functions, its input representation differs from that of the teacher. Reference and query matrices were standardized gene-wise within their respective domains, without sharing gene-wise means or standard deviations between modalities. Standardized values were clipped to
$[-3,3]$.

The SAGA student was optimized using Adam with a learning rate of
$10^{-3}$,
a default batch size of 256 and 30 training epochs.
No weight decay or gradient clipping was applied.

After each epoch, teacher--student top-1 agreement was evaluated on query cells, and the checkpoint with the highest agreement was retained.

\noindent\textbf{Implementation details.}

All models were implemented in PyTorch.
Experiments were repeated using the predefined random seeds described in the benchmark design section. Unless otherwise specified, the same optimization configuration was applied across sections.

\heading{Datasets and preprocessing}
\label{subsec_main_datasets}

\noindent\textbf{Reference and spatial transcriptomics datasets.}

SpCAST integrates labelled single-cell transcriptomic references with unlabelled single-cell-resolution spatial transcriptomics (scST) queries. The evaluated sections include both sequencing-based and imaging-based spatial transcriptomics platforms paired with corresponding single-cell or single-nucleus RNA-seq references. Each analysis uses a reference--query pair consisting of a labelled single-cell reference and an unlabelled spatial transcriptomics query.

For each reference--query pair, cell-type annotations were harmonized into a shared label vocabulary. Original annotations from individual sections were mapped to corresponding harmonized cell-type categories for model training and evaluation.

\noindent\textbf{Expression preprocessing.}

Reference and query expression matrices were processed separately using Scanpy. In the reference datasets, genes expressed in fewer than 3 cells, mitochondrial genes beginning with ``MT-'' or ``mt-'', cells with fewer than 300 total counts and cells with fewer than 600 detected genes were removed. In the spatial query datasets, genes expressed in fewer than 3 spatial cells or spots and mitochondrial genes were removed. Routine cell- or spot-level filtering based on minimum counts or detected genes was not applied to spatial queries in order to preserve the original spatial composition.

When processed expression matrices or technology-specific preprocessing workflows were provided by the original data source, raw-count preprocessing was not repeated.

Raw-count matrices were normalized to 10,000 counts per cell or spatial unit and transformed using
$\log(1+x)$.

Reference and query matrices were subsequently restricted to genes measurable in both modalities.

\noindent\textbf{Input gene selection.}

After restricting the reference and query matrices to genes measurable in both modalities, SpCAST by default used the top 1,500 highly variable genes selected from the shared reference--query gene space as model inputs:

\begin{equation}
\mathcal{G}_{\mathrm{input}}
=
\mathcal{G}_{\mathrm{HVG}}^{(1500)}.
\end{equation}

Here,
$\mathcal{G}_{\mathrm{HVG}}^{(1500)}$
denotes the top 1,500 highly variable genes selected from the shared reference--query gene space.

SpCAST also supports optional marker augmentation, in which the input panel is supplemented with the top 30 Wilcoxon-ranked marker genes for each reference cell type. Marker genes not measurable in both reference and query modalities are excluded. This option was disabled in the default configuration and was used only where explicitly stated.

The final input dimension was
\[
D=|\mathcal{G}_{\mathrm{input}}|.
\]

\noindent\textbf{Platform-specific query preprocessing.}

Because the sequencing-based Slide-tags query used in this study exhibited comparatively sparse single-nucleus expression profiles, we implemented an optional local spatial-expression enhancement procedure for SpCAST, with $\alpha=0.2$ as the default setting. For the head-to-head Slide-tags benchmark reported in the main text, however, this enhancement was disabled ($\alpha=0$) to avoid introducing an additional SpCAST-specific enhancement step that was not applied to the comparison methods. Accordingly, all baseline methods were supplied with the original, unenhanced query expression matrix. We separately evaluated the effect of $\alpha$ in a parameter-sensitivity analysis. The enhancement procedure and the complete parameter-sensitivity analysis are described in the Supplementary Methods and Supplementary Table~\ref{tab:slidetags_enhancement_ablation}.

\heading{Benchmark design and evaluation}
\label{subsec_main_benchmark}

\noindent\textbf{Cell-type transfer benchmark.}

SpCAST was compared with six reference-guided methods: cell2location\cite{Kleshchevnikov2022Cell2location}, SPANN\cite{Yuan2024SPANN}, Spatial-ID\cite{Shen2022SpatialID}, Tangram\cite{Biancalani2021Tangram}, STEM\cite{Hao2024STEM} and SpatialScope\cite{Wan2023SpatialScope}. For each reference--query pair, methods used the same reference and query cells and the same harmonized cell-type vocabulary, and were run on the reference--query shared genes supported by the respective implementation. Query ground-truth labels were used only for final performance evaluation and were not provided during model training or inference.

Baseline methods were run using default or recommended settings from their official implementations, whereas SpCAST used the benchmark configuration described above.

\noindent\textbf{Evaluation metrics.}

For the real-data benchmarks, predicted cell-type labels were compared with the original benchmark annotations using accuracy and weighted F1 score. Cells originally lacking an evaluable annotation were excluded before metric calculation.

Each reference--query pair was evaluated independently using random seeds 0, 42 and 2025. Seed-level results were summarized within each dataset--method combination before dataset-level method comparison.

\noindent\textbf{Dataset-level ranking.}

To compare relative performance across sections, all evaluable methods were ranked independently within each dataset and separately for each evaluation metric, with higher ranks indicating better performance. For a dataset--metric combination containing (M) evaluable methods, the worst-performing method was assigned rank 1 and the best-performing method rank (M). Ties were assigned the average rank. Rank distributions across sections were then used to summarize the relative performance of each method without constructing an additional composite rank score.

\noindent\textbf{Runtime evaluation.}

Real-data runtime was defined as the wall-clock time from the start of model fitting to generation of the final query predictions. For SpCAST, this interval included teacher training, cell-type prediction and decoder output but excluded the subsequent SAGA training used for model interpretation.

The runtime comparison therefore reflects the practical computational cost of completing the primary prediction task under the recommended configuration of each method rather than speed under a strictly matched training budget.

\noindent\textbf{Computational scalability benchmark.}

To evaluate computational scalability under controlled conditions, paired synthetic reference--query datasets were generated. The synthetic spatial query was constructed using the 313-gene panel of the human breast cancer Xenium dataset analysed in this study, thereby representing a targeted single-cell-resolution spatial transcriptomics setting. The reference was fixed at 30,000 cells, 3,000 genes and 10 cell types, whereas the scST query contained 10,000, 100,000, 1,000,000 or 10,000,000 cells.

The synthetic generator included cell-type-specific mean shifts and Poisson sampling noise but no additional reference--query distribution shift. The benchmark therefore retained a Xenium-style targeted gene-panel dimensionality while primarily evaluating computational scalability as the number of query cells increased, rather than annotation difficulty in real tissues.

The scalability benchmark compared SpCAST, cell2location, SPANN, Spatial-ID, SpatialScope, STEM and Tangram. For methods exposing batch-size and epoch controls, the batch size was set to 64 and models were trained for 100 epochs. The reference and spatial-mapping models of cell2location were each trained for 100 epochs. All remaining parameters followed the default or recommended settings of the corresponding official implementations.

All scalability experiments were conducted under the same computational resource limits, with up to 80 GB of GPU memory and 120 GB of system RAM. Runs that failed to complete because of resource exhaustion (for example, exceeding GPU-memory or system-RAM limits) or abnormal termination were recorded as failures. Runtime and memory usage were reported only for successfully completed runs, and no numerical extrapolation was performed for failed runs.

\heading{Cross-species reference transfer}
\label{subsec_main_cross_species}

\noindent\textbf{Human and mouse reference settings.}

Cross-species analyses were performed on the same human prefrontal-cortex Slide-tags query using two reference settings~\cite{Russell2024SlideTagsNature}. The intra-species setting used a human prefrontal-cortex snRNA-seq reference~\cite{Herring2022HumanPFC}, whereas the cross-species setting used a mouse prefrontal-cortex scRNA-seq reference~\cite{Bhattacherjee2019MousePFC}. Mouse genes were mapped to human ortholog symbols before model training. The query dataset, harmonized cell-type vocabulary, model-training procedure and evaluation labels were otherwise held constant between the two settings.

The same feature-selection procedure was applied independently in the human-reference and mouse-reference settings, with the top 1,500 highly variable genes selected from the shared reference–query gene space in each setting. Marker augmentation was not used in these analyses. Because the resulting HVG sets could differ between reference settings, comparisons of SAGA gene-level evidence were restricted to genes retained in both settings.

\noindent\textbf{Cross-species transfer evaluation.}

SpCAST and the cell-type-transfer baseline methods were trained independently under the human-reference and mouse-reference settings and evaluated against the same human query ground-truth labels. Query labels were used only for final evaluation and were not provided during model training or inference. The cross-species performance decrease was calculated as
\[
\Delta\mathrm{Acc}
=
\mathrm{Acc}_{\mathrm{intra}}
-
\mathrm{Acc}_{\mathrm{cross}},
\]
where
\(\mathrm{Acc}_{\mathrm{intra}}\)
and
\(\mathrm{Acc}_{\mathrm{cross}}\)
denote the accuracies obtained using the human and mouse references, respectively.

\noindent\textbf{Gene-level evidence comparison.}

For each shared gene $g$, target cell type $c$ and reference setting
$s\in\{\mathrm{intra},\mathrm{cross}\}$,
the target-versus-other contribution contrast, its positive component and the corresponding strong-evidence threshold were defined as
\begin{equation}
\label{eq:cross_species_evidence}
\begin{aligned}
\Delta_{gc}^{s}
&=
\frac{1}{|\mathcal{I}_{c}^{s}|}
\sum_{i\in\mathcal{I}_{c}^{s}}
\widetilde{\phi}_{gc}^{\,s}
\left(
\tilde{x}_{ig}^{\,s}
\right)
-
\frac{1}{|\overline{\mathcal{I}}_{c}^{s}|}
\sum_{i\in\overline{\mathcal{I}}_{c}^{s}}
\widetilde{\phi}_{gc}^{\,s}
\left(
\tilde{x}_{ig}^{\,s}
\right),
\\[4pt]
E_{gc}^{+,s}
&=
\left[\Delta_{gc}^{s}\right]_{+}
=
\max\left(\Delta_{gc}^{s},0\right),
\\[4pt]
\tau_{c}^{s}
&=
Q_{0.8}
\left(
\left\{
E_{gc}^{+,s}
:
g\in\mathcal{G}_{\mathrm{shared}}
\right\}
\right).
\end{aligned}
\end{equation}

Here,
$\mathcal{I}_{c}^{s}$
denotes the index set of reference cells predicted as cell type $c$ by the teacher model in setting $s$, and
$\overline{\mathcal{I}}_{c}^{s}$
denotes its complement among the reference cells.
The cardinalities
$|\mathcal{I}_{c}^{s}|$
and
$|\overline{\mathcal{I}}_{c}^{s}|$
give the numbers of cells in the two groups, respectively.
The quantity
$\widetilde{\phi}_{gc}^{\,s}(\tilde{x}_{ig}^{\,s})$
is the centred SAGA contribution of gene $g$ to class $c$ for reference cell $i$ in setting $s$.
$\mathcal{G}_{\mathrm{shared}}$
denotes the common gene set used for evidence comparison, and
$Q_{0.8}$
denotes the empirical 80th percentile.

Thus,
$\Delta_{gc}^{s}$
measures the difference in mean centred contribution between teacher-predicted target cells and the remaining reference cells, whereas
$E_{gc}^{+,s}$
retains only the positive class-specific decision evidence. The strong-evidence threshold was determined independently for each cell type and reference setting from the empirical distribution of \(E_{gc}^{+,s}\) across shared genes.

A gene--cell-type pair was operationally classified as species-consistent when
$E_{gc}^{+,\mathrm{intra}}\geq\tau_{c}^{\mathrm{intra}}$
and
$E_{gc}^{+,\mathrm{cross}}\geq\tau_{c}^{\mathrm{cross}}$.
Pairs meeting the corresponding threshold in only one setting were classified as species-inconsistent, whereas pairs meeting neither threshold were classified as weak in both. These categories quantify the retention of strong gene-level evidence across the two reference settings; they do not require similar evidence magnitudes between settings and do not imply that observed differences are attributable exclusively to species.

To further characterize gene-response patterns learned by the independently trained SAGA models, centred response curves for the same gene--cell-type pairs were evaluated on a common standardized-expression grid spanning
$[-3,3]$.

Gene-expression support was evaluated independently in the human and mouse references by comparing each target cell type with all remaining reference cells, providing an expression-level assessment complementary to the SAGA decision evidence.

\heading{Expression reconstruction evaluation}
\label{subsec_main_reconstruction}

\noindent\textbf{Controlled masking benchmark.}

The controlled masking benchmark was designed to quantify recovery of reference-defined cell-type-associated expression signals rather than to reproduce the complete technology-specific dropout process of spatial transcriptomics. Expression reconstruction was evaluated across 53 spatial transcriptomics sections, each paired with its corresponding single-cell or single-nucleus transcriptomic reference. For each section, the reference was restricted to genes measured in the corresponding spatial query. Marker genes were ranked separately for each reference cell type against all remaining reference cells using the Wilcoxon rank-sum test implemented in Scanpy's \texttt{rank\_genes\_groups} function~\cite{Wolf2018Scanpy}, and the top 20 genes per cell type were retained. The union of these genes defined the section-specific reconstruction evaluation panel, $\mathcal{G}_{\mathrm{eval}}$. Spatial query labels were not used for marker selection.

For spatial query expression matrix
$\mathbf{X}^{\mathrm{st}}$,
the masking candidate set was defined as
\begin{equation}
\label{eq:reconstruction_mask}
\Omega
=
\left\{
(i,g):
g\in\mathcal{G}_{\mathrm{eval}},
\ X_{ig}^{\mathrm{st}}>0
\right\}.
\end{equation}
Only originally observed non-zero entries were eligible for masking.

For each spatial section, 10\%, 30\%, 50\% or 70\% of the entries in $\Omega$ were sampled without replacement and set to zero in the model input. Masking was performed globally across eligible cell--gene entries rather than separately within individual cells or genes. Each masking fraction was evaluated using random seeds 0, 7, 42, 123 and 2025. For a given section, masking fraction and random seed, all methods received the same masked query matrix, whereas the corresponding unmasked matrix was retained for evaluation only. For SpCAST, artificially masked entries were excluded from the query-reconstruction loss during training, such that the decoder was optimized only against expression entries that remained visible in the masked query.

\noindent\textbf{Reconstruction methods and ablations.}

The primary benchmark compared SpCAST with SpaGE\cite{SpaGE}, stPlus\cite{stPlus}, Tangram\cite{Biancalani2021Tangram}, SpatialScope\cite{Wan2023SpatialScope} and SpaIM\cite{SpaIM}. Tangram, SpatialScope and SpaIM were run using the official implementations provided by their authors, whereas SpaGE and stPlus were run using implementations integrated into the common benchmarking pipeline. For each section, masking fraction and random seed, all methods received the same masked query matrix and were evaluated at the same masked positions. Reconstructed outputs were aligned to the spatial-query gene order before evaluation.

Two SpCAST ablations were evaluated separately. SpCAST without sc reference removed the single-cell reference, reference-label supervision and reference--query MMD alignment, retaining only the spatial-query reconstruction objective. SpCAST-MLP retained the single-cell reference and the reference-guided training objectives of SpCAST but replaced the KAN encoder and decoder with dimension-matched multilayer perceptrons.

Both ablations used the same input genes, masked query matrices, training schedule and reconstruction-evaluation protocol as SpCAST. The KAN and MLP variants were matched by the input and output dimensions of the corresponding layers, but not by the exact number of trainable parameters. Ablation results were analyzed separately from the primary six-method comparison.

\noindent\textbf{Entry-level reconstruction accuracy.}

Reconstruction accuracy was evaluated exclusively at the artificially masked positions. For each spatial section, masking fraction, random seed and method, expression values at the masked positions were extracted from the unmasked and reconstructed matrices and pooled into matched one-dimensional vectors. Pearson correlation coefficient (PCC) and Spearman rank correlation coefficient were calculated directly between these vectors. The reported values therefore represent section-level pooled correlations across masked cell--gene entries rather than averages of gene-wise or cell-wise correlations. Unmasked entries were excluded from evaluation.

\noindent\textbf{Spatial and decision-evidence validation.}

Cell-type-associated localization of reconstructed marker expression was evaluated in representative human NSCLC CosMx SMI FOV 20. For each target cell type
$c$,
a binary cell-type membership vector
$\mathbf{y}_{c}$
was defined across spatial cells, with entries equal to 1 for cells of type
$c$
and 0 otherwise. For each evaluated marker gene
$g$,
Spearman correlation was calculated between
$\mathbf{y}_{c}$
and the corresponding raw or reconstructed expression vector across cells.

The cell-type-associated localization gain was defined as
\begin{equation}
\label{eq:reconstruction_localization_gain}
\Delta\rho_{g,c}
=
\rho_{\mathrm{S}}
\left(
\widehat{\mathbf{x}}_{g}^{\,\mathrm{rec}},
\mathbf{y}_{c}
\right)
-
\rho_{\mathrm{S}}
\left(
\mathbf{x}_{g}^{\,\mathrm{raw}},
\mathbf{y}_{c}
\right),
\end{equation}
where
$\rho_{\mathrm{S}}$
denotes Spearman correlation,
$\mathbf{x}_{g}^{\,\mathrm{raw}}$
denotes the raw expression vector of gene
$g$
across spatial cells, and
$\widehat{\mathbf{x}}_{g}^{\,\mathrm{rec}}$
denotes the corresponding reconstructed expression vector.
Positive values of
$\Delta\rho_{g,c}$
indicate increased concordance between marker expression and the spatial distribution of the corresponding cell type. Raw and reconstructed maps used identical cell coordinates and the same visualization range within each gene.

To assess reconstruction-associated changes in SAGA marker evidence, two teacher--SAGA models were trained independently using either the measured spatial query expression or the SpCAST-reconstructed query expression, with the same scRNA-seq reference retained in both settings. SAGA marker scores were then calculated on the reference cells for each model. For each cell type, the top $q$ genes ranked by SAGA marker score in the measured-expression setting were selected ($q=20$ by default), and the same gene set was used to summarize marker scores in both settings.

% ============================================================
% Statistical analysis
% ============================================================

\heading{Statistical analysis}
\label{subsec_main_statistics}

For selected marker-gene expression comparisons, the designated target cell type or candidate population was compared with all remaining cells using the Wilcoxon rank-sum test implemented in Scanpy. Reported \(P\) values were unadjusted; these comparisons were used to provide expression-level support for preselected marker genes rather than for genome-wide significance screening.

For paired reconstruction-ablation comparisons, results were first averaged across masking seeds within each benchmark unit to avoid treating repeated random seeds as independent observations. Paired differences between conditions at the benchmark-unit level were then assessed using a two-sided Wilcoxon signed-rank test.

Unless otherwise specified, statistical significance was denoted as *$P<0.05$, **$P<0.01$, ***$P<0.001$ and ****$P<0.0001$; ns, not significant. Statistical tests are specified in the corresponding figure legends where applicable.

\heading{Acknowledgements}

We thank Rhoda E. and Edmund F. Perozzi for editing assistance and constructive feedback on the content and logical flow of the manuscript.

\heading{Author contributions}

Tianzi Jiang and Xiaojuan Sun conceived and supervised the project. Yiyang Zhang and Bokai Zhao implemented the method with assistance from Zongchang Du, Xiaoru Zhang and Minfeng Xu. Yiyang Zhang and Bokai Zhao performed the analyses and prepared the figures. Yiyang Zhang, Bokai Zhao, Xiaojuan Sun and Tianzi Jiang wrote the manuscript. All authors read and approved the final manuscript.

\heading{Competing interests}

% M.X. is an employee of Alibaba Group. The remaining authors declare no
% competing interests.
The authors declare no competing interests.

\heading{Funding}

This work was supported by the Science and Technology Innovation 2030--Brain Science and Brain-Inspired Intelligence Project (grant no. 2021ZD0200200 to T.J.); the National Natural Science Foundation of China (grant no. 62327805 to T.J.); and the National Natural Science Foundation of China (grant no. 12572065 to X.S.).

\heading{Data availability}

All raw datasets analysed in this study are publicly available from the original repositories listed below. The processed AnnData objects used for the analyses are available via \url{https://doi.org/10.5281/zenodo.22171443}.

\textbf{Mouse visual cortex seqFISH+ dataset:}
\url{https://github.com/CaiGroup/seqFISH-PLUS} (scST) and
\href{https://portal.brain-map.org/atlases-and-data/rnaseq/mouse-v1-and-alm-smart-seq}{Allen Institute Mouse V1 and ALM SMART-seq} (scRNA-seq).

\textbf{Mouse hypothalamus MERFISH dataset:}
\url{https://datadryad.org/dataset/doi:10.5061/dryad.8t8s248} (scST) and
\href{https://www.ncbi.nlm.nih.gov/geo/query/acc.cgi?acc=GSE113576}{GEO: GSE113576} (scRNA-seq).

\textbf{Human non-small-cell lung cancer CosMx SMI dataset:}
\url{https://nanostring.com/resources/smi-ffpe-dataset-lung9-rep1-data}(scST) and
\href{https://gbiomed.kuleuven.be/english/cme/research/laboratories/54213024/scRNAseq-NSCLC}{KU Leuven NSCLC scRNA-seq dataset} (scRNA-seq).

\textbf{Human prefrontal cortex Slide-tags dataset:}
\href{https://singlecell.broadinstitute.org/single_cell/study/SCP2167/slide-tags-snrna-seq-on-human-prefrontal-cortex#study-download}{Single Cell Portal: SCP2167} (scST),
\href{https://www.ncbi.nlm.nih.gov/geo/query/acc.cgi?acc=GSE168408}{GEO: GSE168408} (human scRNA-seq) and
\href{https://www.ncbi.nlm.nih.gov/geo/query/acc.cgi?acc=GSE124952}{GEO: GSE124952} (mouse scRNA-seq).

\textbf{Human breast cancer Xenium dataset:}
\url{https://www.10xgenomics.com/datasets/ffpe-human-breast-with-pre-designed-panel-1-standard}(scST) and
\href{https://www.ncbi.nlm.nih.gov/geo/query/acc.cgi?acc=GSE243275}{GEO: GSE243275} (scFFPE-seq reference).

\heading{Code availability}

The source code for SpCAST and the scripts and notebooks used for the analyses reported in this study are publicly available via
\url{https://github.com/sijimochou/SpCAST-1.0.0}.

\end{spacing}
\newpage
% \begin{nolinenumbers}
\Heading{References}
\bibliographystyle{nature}
\bibliography{main}

@article{Chen2018TissuesCellTypes,
  title={From tissues to cell types and back: single-cell gene expression analysis of tissue architecture},
  author={Chen, Xi and Teichmann, Sarah A and Meyer, Kerstin B},
  journal={Annual Review of Biomedical Data Science},
  volume={1},
  number={1},
  pages={29--51},
  year={2018},
  publisher={Annual Reviews}
}

@article{TabulaSapiens2022,
  title={The Tabula Sapiens: A multiple-organ, single-cell transcriptomic atlas of humans},
  author={{Tabula Sapiens Consortium} and Jones, Robert C and Karkanias, Jim and others},
  journal={Science},
  volume={376},
  number={6594},
  pages={eabl4896},
  year={2022},
  publisher={American Association for the Advancement of Science}
}

@article{Stahl2016SpatialTranscriptomics,
  title={Visualization and analysis of gene expression in tissue sections by spatial transcriptomics},
  author={St{\aa}hl, Patrik L and Salm{\'e}n, Fredrik and Vickovic, Sanja and others},
  journal={Science},
  volume={353},
  number={6294},
  pages={78--82},
  year={2016},
  publisher={American Association for the Advancement of Science}
}

@article{Chen2015MERFISH,
  title={Spatially resolved, highly multiplexed RNA profiling in single cells},
  author={Chen, Kok Hao and Boettiger, Alistair N and Moffitt, Jeffrey R and others},
  journal={Science},
  volume={348},
  number={6233},
  pages={aaa6090},
  year={2015},
  publisher={American Association for the Advancement of Science}
}

@article{Lubeck2014SeqFISH,
  title={Single-cell in situ RNA profiling by sequential hybridization},
  author={Lubeck, Eric and Coskun, Ahmet F and Zhiyentayev, Timur and others},
  journal={Nature methods},
  volume={11},
  number={4},
  pages={360--361},
  year={2014},
  publisher={Nature Publishing Group US New York}
}

@article{Eng2019SeqFISHPlus,
  title={Transcriptome-scale super-resolved imaging in tissues by RNA seqFISH+},
  author={Eng, Chee-Huat Linus and Lawson, Michael and Zhu, Qian and others},
  journal={Nature},
  volume={568},
  number={7751},
  pages={235--239},
  year={2019},
  publisher={Nature Publishing Group UK London}
}

@article{Codeluppi2018OsmFISH,
  title={Spatial organization of the somatosensory cortex revealed by osmFISH},
  author={Codeluppi, Simone and Borm, Lars E and Zeisel, Amit and others},
  journal={Nature methods},
  volume={15},
  number={11},
  pages={932--935},
  year={2018},
  publisher={Nature Publishing Group US New York}
}

@article{Chen2022StereoSeqOrganogenesis,
  title={Spatiotemporal transcriptomic atlas of mouse organogenesis using DNA nanoball-patterned arrays},
  author={Chen, Ao and Liao, Sha and Cheng, Mengnan and others},
  journal={Cell},
  volume={185},
  number={10},
  pages={1777--1792},
  year={2022},
  publisher={Elsevier}
}

@article{Russell2024SlideTagsNature,
  title={Slide-tags enables single-nucleus barcoding for multimodal spatial genomics},
  author={Russell, Andrew JC and Weir, Jackson A and Nadaf, Naeem M and others},
  journal={Nature},
  volume={625},
  number={7993},
  pages={101--109},
  year={2024},
  publisher={Nature Publishing Group UK London}
}

@article{He2022SpatialMolecularImaging,
  title={High-plex imaging of RNA and proteins at subcellular resolution in fixed tissue by spatial molecular imaging},
  author={He, Shanshan and Bhatt, Ruchir and Brown, Carl and others},
  journal={Nature biotechnology},
  volume={40},
  number={12},
  pages={1794--1806},
  year={2022},
  publisher={Nature Publishing Group US New York}
}

@article{Janesick2023Xenium,
  title={High resolution mapping of the tumor microenvironment using integrated single-cell, spatial and in situ analysis},
  author={Janesick, Amanda and Shelansky, Robert and Gottscho, Andrew D and others},
  journal={Nature communications},
  volume={14},
  number={1},
  pages={8353},
  year={2023},
  publisher={Nature Publishing Group UK London}
}

@article{Fan2023SPASCER,
  title={SPASCER: spatial transcriptomics annotation at single-cell resolution},
  author={Fan, Zhiwei and Luo, Yangyang and Lu, Huifen and others},
  journal={Nucleic Acids Research},
  volume={51},
  number={D1},
  pages={D1138--D1149},
  year={2023},
  publisher={Oxford University Press}
}

@article{Shen2022SpatialID,
  title={Spatial-ID: a cell typing method for spatially resolved transcriptomics via transfer learning and spatial embedding},
  author={Shen, Rongbo and Liu, Lin and Wu, Zihan and others},
  journal={Nature communications},
  volume={13},
  number={1},
  pages={7640},
  year={2022},
  publisher={Nature Publishing Group UK London}
}

@article{Hao2024STEM,
  title={STEM enables mapping of single-cell and spatial transcriptomics data with transfer learning},
  author={Hao, Minsheng and Luo, Erpai and Chen, Yixin and others},
  journal={Communications biology},
  volume={7},
  number={1},
  pages={56},
  year={2024},
  publisher={Nature Publishing Group UK London}
}

@article{Biancalani2021Tangram,
  title={Deep learning and alignment of spatially resolved single-cell transcriptomes with Tangram},
  author={Biancalani, Tommaso and Scalia, Gabriele and Buffoni, Lorenzo and others},
  journal={Nature methods},
  volume={18},
  number={11},
  pages={1352--1362},
  year={2021},
  publisher={Nature Publishing Group US New York}
}

@article{Kleshchevnikov2022Cell2location,
  title={Cell2location maps fine-grained cell types in spatial transcriptomics},
  author={Kleshchevnikov, Vitalii and Shmatko, Artem and Dann, Emma and others},
  journal={Nature biotechnology},
  volume={40},
  number={5},
  pages={661--671},
  year={2022},
  publisher={Nature Publishing Group US New York}
}

@article{Wan2023SpatialScope,
  title={Integrating spatial and single-cell transcriptomics data using deep generative models with SpatialScope},
  author={Wan, Xiaomeng and Xiao, Jiashun and Tam, Sindy Sing Ting and others},
  journal={Nature Communications},
  volume={14},
  number={1},
  pages={7848},
  year={2023},
  publisher={Nature Publishing Group UK London}
}

@article{Yuan2024SPANN,
  title={SPANN: annotating single-cell resolution spatial transcriptome data with scRNA-seq data},
  author={Yuan, Musu and Wan, Hui and Wang, Zihao and others},
  journal={Briefings in Bioinformatics},
  volume={25},
  number={2},
  pages={bbad533},
  year={2024},
  publisher={Oxford University Press}
}

@article{Li2022BenchmarkingIntegration,
  title={Benchmarking spatial and single-cell transcriptomics integration methods for transcript distribution prediction and cell type deconvolution},
  author={Li, Bin and Zhang, Wen and Guo, Chuang and Xu, Hao and Li, Longfei and Fang, Minghao and Hu, Yinlei and Zhang, Xinye and Yao, Xinfeng and Tang, Meifang and others},
  journal={Nature methods},
  volume={19},
  number={6},
  pages={662--670},
  year={2022},
  publisher={Nature Publishing Group US New York}
}

@article{SpaGE,
  title={SpaGE: spatial gene enhancement using scRNA-seq},
  author={Abdelaal, Tamim and Mourragui, Soufiane and Mahfouz, Ahmed and Reinders, Marcel JT},
  journal={Nucleic acids research},
  volume={48},
  number={18},
  pages={e107--e107},
  year={2020},
  publisher={Oxford University Press}
}

@article{stPlus,
  title={stPlus: a reference-based method for the accurate enhancement of spatial transcriptomics},
    author={Chen, Shengquan and Zhang, Boheng and Chen, Xiaoyang and Zhang, Xuegong and Jiang, Rui},
  journal={Bioinformatics},
  volume={37},
  number={Supplement\_1},
  pages={i299--i307},
  year={2021},
  publisher={Oxford University Press}
}

@article{SpaIM,
  title={SpaIM: single-cell spatial transcriptomics imputation via style transfer},
  author={Li, Bo and Tang, Ziyang and Budhkar, Aishwarya and Liu, Xiang and Zhang, Tonglin and Yang, Baijian and Su, Jing and Song, Qianqian},
  journal={Nature Communications},
  volume={16},
  number={1},
  pages={7861},
  year={2025},
  publisher={Nature Publishing Group UK London}
}

@inproceedings{Liu2024KAN,
  title={{KAN}: Kolmogorov--Arnold Networks},
  author={Liu, Ziming and Wang, Yixuan and Vaidya, Sachin and Ruehle, Fabian and Halverson, James and Soljacic, Marin and Hou, Thomas Y. and Tegmark, Max},
  booktitle={International Conference on Learning Representations},
  year={2025}
}

@inproceedings{LundbergLee2017SHAP,
  title={A Unified Approach to Interpreting Model Predictions},
  author={Lundberg, Scott M and Lee, Su-In},
  booktitle={Advances in Neural Information Processing Systems},
  volume={30},
  pages={4765--4774},
  year={2017}
}

@article{HastieTibshirani1987GAM,
  title={Generalized additive models: some applications},
  author={Hastie, Trevor and Tibshirani, Robert},
  journal={Journal of the American Statistical Association},
  volume={82},
  number={398},
  pages={371--386},
  year={1987},
  publisher={Taylor \& Francis}
}

@article{Gretton2012MMD,
  title={A kernel two-sample test},
  author={Gretton, Arthur and Borgwardt, Karsten M and Rasch, Malte J and others},
  journal={The journal of machine learning research},
  volume={13},
  number={1},
  pages={723--773},
  year={2012},
  publisher={JMLR. org}
}

@article{Hinton2015Distilling,
  title={Distilling the Knowledge in a Neural Network},
  author={Hinton, Geoffrey and Vinyals, Oriol and Dean, Jeff},
  journal={arXiv preprint arXiv:1503.02531},
  year={2015}
}

@article{Agarwal2021NAM,
  title={Neural Additive Models: Interpretable Machine Learning with Neural Nets},
  author={Agarwal, Rishabh and Melnick, Levi and Frosst, Nicholas and Zhang, Xuezhou and Lengerich, Ben and Caruana, Rich and Hinton, Geoffrey E.},
  journal={Advances in Neural Information Processing Systems},
  volume={34},
  pages={4699--4711},
  year={2021}
}

@inproceedings{LopezPaz2016Generalized,
  title={Unifying Distillation and Privileged Information},
  author={Lopez-Paz, David and Bottou, L{\'e}on and Sch{\"o}lkopf, Bernhard and Vapnik, Vladimir},
  booktitle={International Conference on Learning Representations},
  year={2016}
}

@article{chen2023transformer,
title={Transformer for one stop interpretable cell type annotation},
author={Chen, Jiawei and Xu, Hao and Tao, Wanyu and Chen, Zhaoxiong and Zhao, Yuxuan and Han, Jing-Dong J},
journal={Nature Communications},
volume={14},
number={1},
pages={223},
year={2023},
publisher={Nature Publishing Group UK London}
}

@article{zhong2024interpretable,
title={Interpretable spatially aware dimension reduction of spatial transcriptomics with STAMP},
author={Zhong, Chengwei and Ang, Kok Siong and Chen, Jinmiao},
journal={Nature Methods},
volume={21},
number={11},
pages={2072--2083},
year={2024},
publisher={Nature Publishing Group US New York}
}

@article{chitra2025mapping,
title={Mapping the topography of spatial gene expression with interpretable deep learning},
author={Chitra, Uthsav and Arnold, Brian J and Sarkar, Hirak and Sanno, Kohei and Ma, Cong and Lopez-Darwin, Sereno and Raphael, Benjamin J},
journal={Nature Methods},
volume={22},
number={2},
pages={298--309},
year={2025},
publisher={Nature Publishing Group US New York}
}

@article{gold2026scoring,
  title={Scoring gene importance by interpreting single-cell foundation models},
  author={Gold, Maxwell P and Reyes, Miguel and Diamant, Nathaniel and Kuo, Tony and Hajiramezanali, Ehsan and Newburger, Jane W and Son, Mary Beth F and Lee, Pui Y and Scalia, Gabriele and BenTaieb, Aicha and others},
  journal={Nature Biotechnology},
  year={2026},
  doi={10.1038/s41587-026-03112-5}
}

@article{Wolf2018Scanpy,
  title={SCANPY: large-scale single-cell gene expression data analysis},
  author={Wolf, F Alexander and Angerer, Philipp and Theis, Fabian J},
  journal={Genome biology},
  volume={19},
  number={1},
  pages={15},
  year={2018},
  publisher={Springer}
}

@article{Moffitt2018Hypothalamus,
  title={Molecular, spatial, and functional single-cell profiling of the hypothalamic preoptic region},
  author={Moffitt, Jeffrey R and Bambah-Mukku, Dhananjay and Eichhorn, Stephen W and others},
  journal={Science},
  volume={362},
  number={6416},
  pages={eaau5324},
  year={2018},
  publisher={American Association for the Advancement of Science}
}

@article{Tasic2016MouseV1Taxonomy,
  title={Adult mouse cortical cell taxonomy revealed by single cell transcriptomics},
  author={Tasic, Bosiljka and Menon, Vilas and Nguyen, Thuc Nghi and others},
  journal={Nature neuroscience},
  volume={19},
  number={2},
  pages={335--346},
  year={2016},
  publisher={Nature Publishing Group US New York}
}

@article{Tasic2018NeocorticalAreas,
  title={Shared and distinct transcriptomic cell types across neocortical areas},
  author={Tasic, Bosiljka and Yao, Zizhen and Graybuck, Lucas T and others},
  journal={Nature},
  volume={563},
  number={7729},
  pages={72--78},
  year={2018},
  publisher={Nature Publishing Group UK London}
}

@article{Yao2021IsocortexHPF,
  title={A taxonomy of transcriptomic cell types across the isocortex and hippocampal formation},
  author={Yao, Zizhen and Van Velthoven, Cindy TJ and Nguyen, Thuc Nghi and others},
  journal={Cell},
  volume={184},
  number={12},
  pages={3222--3241},
  year={2021},
  publisher={Elsevier}
}

@article{Yao2023WholeMouseBrainAtlas,
  title={A high-resolution transcriptomic and spatial atlas of cell types in the whole mouse brain},
  author={Yao, Zizhen and Van Velthoven, Cindy TJ and Kunst, Michael and others},
  journal={Nature},
  volume={624},
  number={7991},
  pages={317--332},
  year={2023},
  publisher={Nature Publishing Group UK London}
}

@article{Zhang2023WholeMouseBrainMERFISH,
  title={Molecularly defined and spatially resolved cell atlas of the whole mouse brain},
  author={Zhang, Meng and Pan, Xingjie and Jung, Won and others},
  journal={Nature},
  volume={624},
  number={7991},
  pages={343--354},
  year={2023},
  publisher={Nature Publishing Group UK London}
}

@article{Langlieb2023MouseBrainCytoarchitecture,
  title={The molecular cytoarchitecture of the adult mouse brain},
  author={Langlieb, Jonah and Sachdev, Nina S and Balderrama, Karol S and others},
  journal={Nature},
  volume={624},
  number={7991},
  pages={333--342},
  year={2023},
  publisher={Nature Publishing Group UK London}
}

@article{Zeisel2018MouseNervousSystem,
  title={Molecular architecture of the mouse nervous system},
  author={Zeisel, Amit and Hochgerner, Hannah and L{\"o}nnerberg, Peter and others},
  journal={Cell},
  volume={174},
  number={4},
  pages={999--1014},
  year={2018},
  publisher={Elsevier}
}

@article{Saunders2018DropVizMouseBrain,
  title={Molecular diversity and specializations among the cells of the adult mouse brain},
  author={Saunders, Arpiar and Macosko, Evan Z and Wysoker, Alec and others},
  journal={Cell},
  volume={174},
  number={4},
  pages={1015--1030},
  year={2018},
  publisher={Elsevier}
}

@article{Campbell2017HypothalamusCensus,
  title={A molecular census of arcuate hypothalamus and median eminence cell types},
  author={Campbell, John N and Macosko, Evan Z and Fenselau, Henning and others},
  journal={Nature neuroscience},
  volume={20},
  number={3},
  pages={484--496},
  year={2017},
  publisher={Nature Publishing Group US New York}
}

@article{Belgard2011NeocorticalLayers,
  title={A transcriptomic atlas of mouse neocortical layers},
  author={Belgard, T Grant and Marques, Ana C and Oliver, Peter L and others},
  journal={Neuron},
  volume={71},
  number={4},
  pages={605--616},
  year={2011},
  publisher={Elsevier}
}

@article{Cahoy2008GlialTranscriptome,
  title={A transcriptome database for astrocytes, neurons, and oligodendrocytes: a new resource for understanding brain development and function},
  author={Cahoy, John D and Emery, Ben and Kaushal, Amit and others},
  journal={Journal of Neuroscience},
  volume={28},
  number={1},
  pages={264--278},
  year={2008},
  publisher={Society for Neuroscience}
}

@article{Siletti2023HumanBrainCellTypes,
  title={Transcriptomic diversity of cell types across the adult human brain},
  author={Siletti, Kimberly and Hodge, Rebecca and Mossi Albiach, Alejandro and others},
  journal={Science},
  volume={382},
  number={6667},
  pages={eadd7046},
  year={2023},
  publisher={American Association for the Advancement of Science}
}

@article{Herring2022HumanPFC,
  title={Human prefrontal cortex gene regulatory dynamics from gestation to adulthood at single-cell resolution},
  author={Herring, Charles A and Simmons, Rebecca K and Freytag, Saskia and others},
  journal={Cell},
  volume={185},
  number={23},
  pages={4428--4447},
  year={2022},
  publisher={Elsevier}
}

@article{Bhattacherjee2019MousePFC,
  title={Cell type-specific transcriptional programs in mouse prefrontal cortex during adolescence and addiction},
  author={Bhattacherjee, Aritra and Djekidel, Mohamed Nadhir and Chen, Renchao and others},
  journal={Nature communications},
  volume={10},
  number={1},
  pages={4169},
  year={2019},
  publisher={Nature Publishing Group UK London}
}

@article{Molyneaux2007CorticalSubtype,
  title={Neuronal subtype specification in the cerebral cortex},
  author={Molyneaux, Bradley J. and Arlotta, Paola and Menezes, Jo{\~a}o R. L. and others},
  journal={Nature Reviews Neuroscience},
  volume={8},
  number={6},
  pages={427--437},
  year={2007},
  doi={10.1038/nrn2151}
}

@article{Nieto2004CuxGenes,
  title={Expression of Cux-1 and Cux-2 in the subventricular zone and upper layers II--IV of the cerebral cortex},
  author={Nieto, Marta and Monuki, Edwin S. and Tang, Hua and others},
  journal={Journal of Comparative Neurology},
  volume={479},
  number={2},
  pages={168--180},
  year={2004},
  doi={10.1002/cne.20322}
}

@article{Cubelos2010Cux1Cux2,
  title={Cux1 and Cux2 regulate dendritic branching, spine morphology, and synapses of the upper layer neurons of the cortex},
  author={Cubelos, Beatriz and Sebasti{\'a}n-Serrano, {\'A}ngel and Beccari, Leonardo and others},
  journal={Neuron},
  volume={66},
  number={4},
  pages={523--535},
  year={2010},
  doi={10.1016/j.neuron.2010.04.038}
}

@article{Clark2020RORB,
  title={Cortical ROR{\ensuremath{\beta}} is required for layer 4 transcriptional identity and barrel integrity},
  author={Clark, Erin A. and Rutlin, Michael and Capano, Lucia and others},
  journal={eLife},
  volume={9},
  pages={e52370},
  year={2020},
  doi={10.7554/eLife.52370}
}

@article{Miskic2021CUX2,
  title={Adult upper cortical layer specific transcription factor CUX2 is expressed in transient subplate and marginal zone neurons of the developing human brain},
  author={Mi{\v{s}}ki{\'c}, Terezija and Kostovi{\'c}, Ivica and Rasin, Mladen-Roko and others},
  journal={Cells},
  volume={10},
  number={2},
  pages={415},
  year={2021},
  doi={10.3390/cells10020415}
}

@article{Kast2019FOXP2,
  title={FOXP2 exhibits projection neuron class specific expression, but is not required for multiple aspects of cortical histogenesis},
  author={Kast, Ryan J. and Lanjewar, Alexandra L. and Smith, Colton D. and others},
  journal={eLife},
  volume={8},
  pages={e42012},
  year={2019},
  doi={10.7554/eLife.42012}
}

@article{Romero2012GeneExpressionEvolution,
  title={Comparative studies of gene expression and the evolution of gene regulation},
  author={Romero, Irene Gallego and Ruvinsky, Ilya and Gilad, Yoav},
  journal={Nature Reviews Genetics},
  volume={13},
  number={7},
  pages={505--516},
  year={2012},
  publisher={Nature Publishing Group UK London}
}

@article{Bakken2021ComparativeMotorCortex,
  title={Comparative cellular analysis of motor cortex in human, marmoset and mouse},
  author={Bakken, Trygve E and Jorstad, Nikolas L and Hu, Qiwen and Lake, Blue B and Tian, Wei and Kalmbach, Brian E and Crow, Megan and Hodge, Rebecca D and Krienen, Fenna M and Sorensen, Staci A and others},
  journal={Nature},
  volume={598},
  number={7879},
  pages={111--119},
  year={2021},
  publisher={Nature Publishing Group UK London}
}

@article{Fang2022MERFISHHumanMouse,
  title={Conservation and divergence of cortical cell organization in human and mouse revealed by MERFISH},
  author={Fang, Rongxin and Xia, Chenglong and Close, Jennie L and Zhang, Meng and He, Jiang and Huang, Zhengkai and Halpern, Aaron R and Long, Brian and Miller, Jeremy A and Lein, Ed S and others},
  journal={Science},
  volume={377},
  number={6601},
  pages={56--62},
  year={2022},
  publisher={American Association for the Advancement of Science}
}

@article{Jorstad2023ComparativeTranscriptomics,
  title={Comparative transcriptomics reveals human-specific cortical features},
  author={Jorstad, Nikolas L and Song, Janet HT and Exposito-Alonso, David and Suresh, Hamsini and Castro-Pacheco, Nathan and Krienen, Fenna M and Yanny, Anna Marie and Close, Jennie and Gelfand, Emily and Long, Brian and others},
  journal={Science},
  volume={382},
  number={6667},
  pages={eade9516},
  year={2023},
  publisher={American Association for the Advancement of Science}
}

@article{Krienen2020PrimateInterneurons,
  title={Innovations present in the primate interneuron repertoire},
  author={Krienen, Fenna M and Goldman, Melissa and Zhang, Qiangge and others},
  journal={Nature},
  volume={586},
  number={7828},
  pages={262--269},
  year={2020},
  publisher={Nature Publishing Group UK London}
}

@article{hodge2019conserved,
  title={Conserved cell types with divergent features in human versus mouse cortex},
  author={Hodge, Rebecca D and Bakken, Trygve E and Miller, Jeremy A and Smith, Kimberly A and Barkan, Eliza R and Graybuck, Lucas T and Close, Jennie L and Long, Brian and Johansen, Nelson and Penn, Osnat and others},
  journal={Nature},
  volume={573},
  number={7772},
  pages={61--68},
  year={2019},
  publisher={Nature Publishing Group UK London}
}

@article{Wu2021BreastCancerAtlas,
  title={A single-cell and spatially resolved atlas of human breast cancers},
  author={Wu, Sunny Z and Al-Eryani, Ghamdan and Roden, Daniel Lee and others},
  journal={Nature genetics},
  volume={53},
  number={9},
  pages={1334--1347},
  year={2021},
  publisher={Nature Publishing Group US New York}
}

@article{Kumar2023HumanBreastAtlas,
  title={A spatially resolved single-cell genomic atlas of the adult human breast},
  author={Kumar, Tapsi and Nee, Kevin and Wei, Runmin and others},
  journal={Nature},
  volume={620},
  number={7972},
  pages={181--191},
  year={2023},
  publisher={Nature Publishing Group UK London}
}

@article{Reed2024HumanBreastCellAtlas,
  title={A single-cell atlas enables mapping of homeostatic cellular shifts in the adult human breast},
  author={Reed, Austin D and Pensa, Sara and Steif, Adi and others},
  journal={Nature genetics},
  volume={56},
  number={4},
  pages={652--662},
  year={2024},
  publisher={Nature Publishing Group US New York}
}

@article{Gray2022HumanBreastAtlas,
  title={A human breast atlas integrating single-cell proteomics and transcriptomics},
  author={Gray, G Kenneth and Li, Carman Man-Chung and Rosenbluth, Jennifer M and others},
  journal={Developmental cell},
  volume={57},
  number={11},
  pages={1400--1420},
  year={2022},
  publisher={Elsevier}
}

@article{Azizi2018BreastImmuneMap,
  title={Single-cell map of diverse immune phenotypes in the breast tumor microenvironment},
  author={Azizi, Elham and Carr, Ambrose J and Plitas, George and others},
  journal={Cell},
  volume={174},
  number={5},
  pages={1293--1308},
  year={2018},
  publisher={Elsevier}
}

@article{Jackson2020BreastPathologyLandscape,
  title={The single-cell pathology landscape of breast cancer},
  author={Jackson, Hartland W and Fischer, Jana R and Zanotelli, Vito RT and others},
  journal={Nature},
  volume={578},
  number={7796},
  pages={615--620},
  year={2020},
  publisher={Nature Publishing Group UK London}
}

@article{Nguyen2018BreastEpithelial,
  title={Profiling human breast epithelial cells using single cell RNA sequencing identifies cell diversity},
  author={Nguyen, Quy H and Pervolarakis, Nicholas and Blake, Kerrigan and others},
  journal={Nature communications},
  volume={9},
  number={1},
  pages={2028},
  year={2018},
  publisher={Nature Publishing Group UK London}
}

@article{BhatNakshatri2021HealthyBreast,
  title={A single-cell atlas of the healthy breast tissues reveals clinically relevant clusters of breast epithelial cells},
  author={Bhat-Nakshatri, Poornima and Gao, Hongyu and Sheng, Liu and others},
  journal={Cell Reports Medicine},
  volume={2},
  number={3},
  year={2021},
  publisher={Elsevier}
}

@article{Xu2024BreastTumorAtlas,
  title={A comprehensive single-cell breast tumor atlas defines epithelial and immune heterogeneity and interactions predicting anti-PD-1 therapy response},
  author={Xu, Lily and Saunders, Kaitlyn and Huang, Shao-Po and others},
  journal={Cell Reports Medicine},
  volume={5},
  number={5},
  year={2024},
  publisher={Elsevier}
}

@article{Yang2024PanCancerBCells,
  title={Pan-cancer single-cell dissection reveals phenotypically distinct B cell subtypes},
  author={Yang, Yu and Chen, Xueyan and Pan, Jieying and others},
  journal={Cell},
  volume={187},
  number={17},
  pages={4790--4811},
  year={2024},
  publisher={Elsevier}
}

@article{Lambrechts2018LungTME,
  title={Phenotype molding of stromal cells in the lung tumor microenvironment},
  author={Lambrechts, Diether and Wauters, Els and Boeckx, Bram and others},
  journal={Nature medicine},
  volume={24},
  number={8},
  pages={1277--1289},
  year={2018},
  publisher={Nature Publishing Group US New York}
}

@article{DeZuani2024NSCLCSpatialSingleCell,
  title={Single-cell and spatial transcriptomics analysis of non-small cell lung cancer},
  author={De Zuani, Marco and Xue, Haoliang and Park, Jun Sung and others},
  journal={Nature communications},
  volume={15},
  number={1},
  pages={4388},
  year={2024},
  publisher={Nature Publishing Group UK London}
}

@article{Luo2022PanCancerCAF,
  title={Pan-cancer single-cell analysis reveals the heterogeneity and plasticity of cancer-associated fibroblasts in the tumor microenvironment},
  author={Luo, Han and Xia, Xuyang and Huang, Li-Bin and others},
  journal={Nature communications},
  volume={13},
  number={1},
  pages={6619},
  year={2022},
  publisher={Nature Publishing Group UK London}
}

@article{CellMarker2023,
  title={CellMarker 2.0: an updated database of manually curated cell markers in human/mouse and web tools based on scRNA-seq data},
  author={Hu, Congxue and Li, Tengyue and Xu, Yingqi and others},
  journal={Nucleic acids research},
  volume={51},
  number={D1},
  pages={D870--D876},
  year={2023},
  publisher={Oxford University Press}
}

@article{PanglaoDB2019,
  title={PanglaoDB: a web server for exploration of mouse and human single-cell RNA sequencing data},
  author={Franz{\'e}n, Oscar and Gan, Li-Ming and Bj{\"o}rkegren, Johan LM},
  journal={Database},
  volume={2019},
  pages={baz046},
  year={2019},
  publisher={Oxford University Press}
}

@article{CellMarker2019,
  title={CellMarker: a manually curated resource of cell markers in human and mouse},
  author={Zhang, Xinxin and Lan, Yujia and Xu, Jinyuan and others},
  journal={Nucleic acids research},
  volume={47},
  number={D1},
  pages={D721--D728},
  year={2019},
  publisher={Oxford University Press}
}
% \end{nolinenumbers}

\newpage
% \setcounter{figure}{0}
% \renewcommand{\figurename}{\textbf{Extended Data Figure}}
% \renewcommand{\thefigure}{}

% \setcounter{table}{0}
% \renewcommand{\tablename}{\textbf{Extended Data Table}}
% \renewcommand{\thetable}{S\arabic{table}}

% ============================================================
% Extended Data Figures
% ============================================================

\clearpage

% \section*{Extended Data}
\setcounter{extendedfigure}{0}

% ============================================================
% Extended Data Figure 1
% ============================================================

\begin{extendedfigure}[p]
    \centering
    \includegraphics[
        width=\textwidth,
        % height=0.76\textheight,
        keepaspectratio
    ]{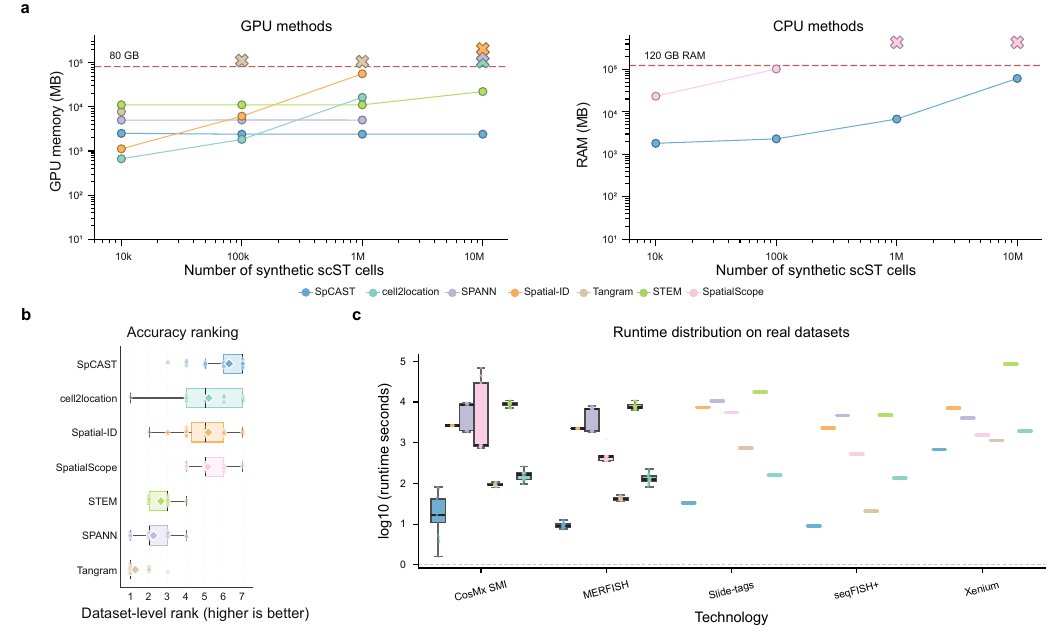}
    \caption{
    \textbf{Computational resource use, accuracy ranking and runtime benchmarks.}
    \textbf{a,} Peak memory use on synthetic scST queries containing between 10{,}000 and 10 million cells. Left, peak GPU memory for GPU-based methods; right, peak system memory for CPU-based methods. Lines connect measurements from the same method. Dashed lines indicate the available 80~GB of GPU memory and 120~GB of system memory, and crosses denote runs that exceeded the corresponding resource limit.
    \textbf{b,} Cross-dataset rank scores based on accuracy. Accuracy was averaged across three random seeds within each dataset before the seven methods were ranked; higher scores indicate better aggregate performance.
    \textbf{c,} Runtime distributions on real datasets grouped by spatial transcriptomic technology. Colors denote methods, and the vertical axis shows $\log_{10}$-transformed runtime in seconds.
    }
    \label{fig:benchmark_resources}
\end{extendedfigure}

\clearpage

% ============================================================
% Extended Data Figure 2
% ============================================================

\begin{extendedfigure}[p]
    \centering
    \includegraphics[
        width=\textwidth,
        % height=0.76\textheight,
        keepaspectratio
    ]{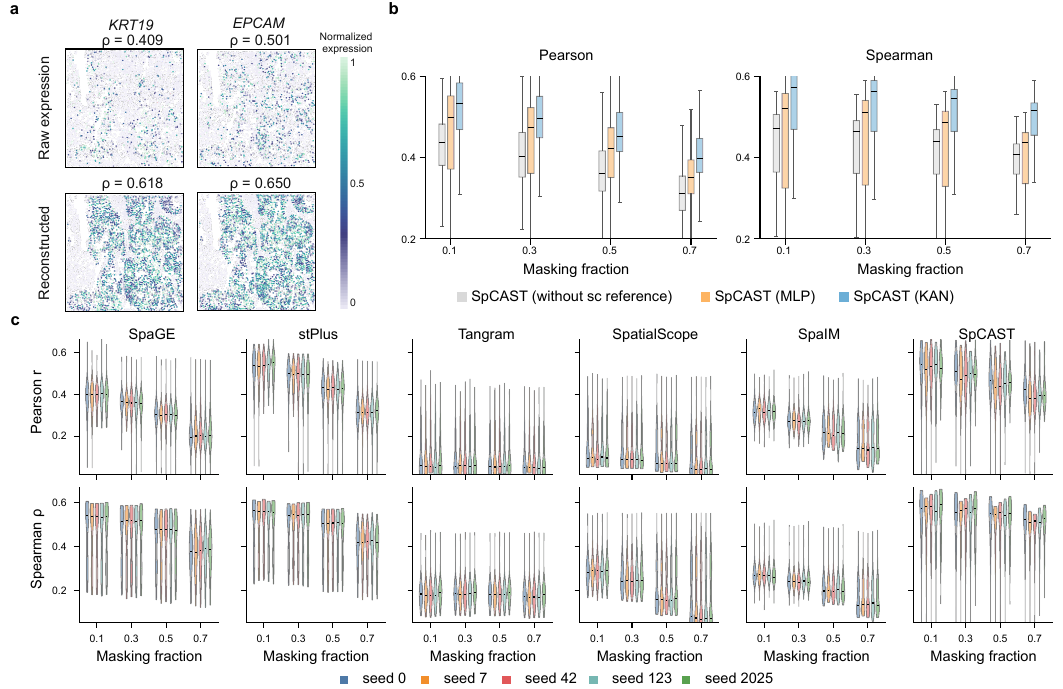}
    \caption{
    \textbf{Additional spatial examples, model ablations and masking-seed stability for expression reconstruction.}
    \textbf{a,} Measured and SpCAST-reconstructed expression of \textit{KRT19} and \textit{EPCAM} in representative CosMx SMI FOV 20 shown in Fig.~\ref{fig:reconstruction}. Spearman correlations with the corresponding tumour-cell spatial reference are shown above each map; expression intensities are normalized independently within each gene.
    \textbf{b,} Pearson and Spearman correlations for SpCAST without the scRNA-seq reference, SpCAST with the KAN components replaced by multilayer perceptrons and the complete SpCAST(KAN) model. Each observation represents one dataset--masking-seed record ($n=265$ observations per variant and masking fraction). Centre lines indicate medians, boxes the interquartile range and whiskers values within 1.5 times the interquartile range.
    \textbf{c,} Pearson and Spearman correlation distributions stratified by masking seed for the six reconstruction methods and four masking fractions. Each distribution contains one observation per dataset ($n=53$ datasets per method, seed and masking fraction).
    }
    \label{fig:reconstruction_extended}
\end{extendedfigure}

\clearpage

% ============================================================
% Extended Data Figure 3
% ============================================================

\begin{extendedfigure}[p]
    \centering
    \includegraphics[
        width=\textwidth,
        % height=0.76\textheight,
        keepaspectratio
    ]{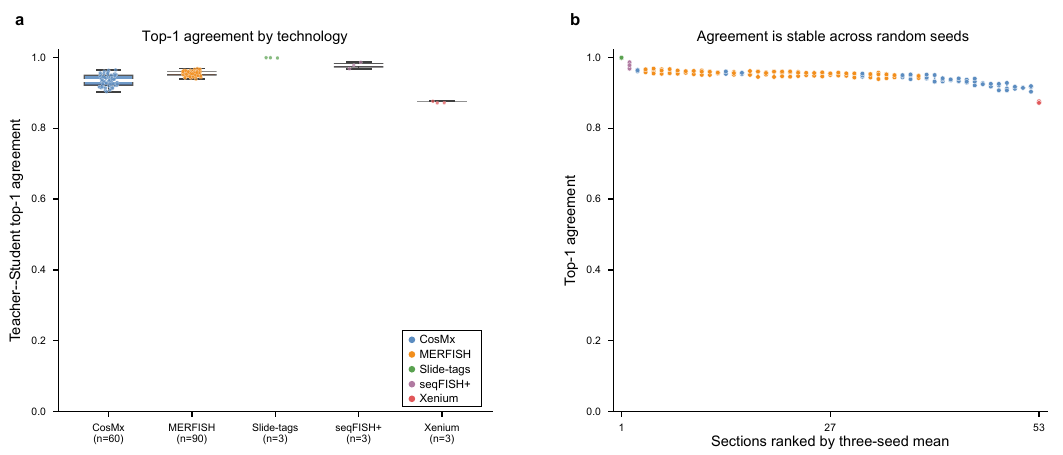}
    \caption{
    \textbf{Teacher--student top-1 agreement of SAGA across the spatial transcriptomics benchmark.}
    \textbf{a,} Teacher--student top-1 agreement across 159 section--seed runs from 53 benchmark spatial transcriptomic sections, grouped by spatial transcriptomics technology (CosMx, \(n=60\); MERFISH, \(n=90\); Slide-tags, \(n=3\); seqFISH+, \(n=3\); Xenium, \(n=3\)). Each point represents one section--seed run.
    \textbf{b,} Section-level teacher--student top-1 agreement for the three predefined random seeds (0, 42 and 2025) across the 53 benchmark sections. Sections are ordered by their mean agreement across the three seeds.
    }
    \label{fig:saga_fidelity_extended}
\end{extendedfigure}

\clearpage

% ============================================================
% Extended Data Figure 4
% ============================================================

\begin{extendedfigure}[p]
    \centering
    \includegraphics[
        width=\textwidth,
        % height=0.76\textheight,
        keepaspectratio
    ]{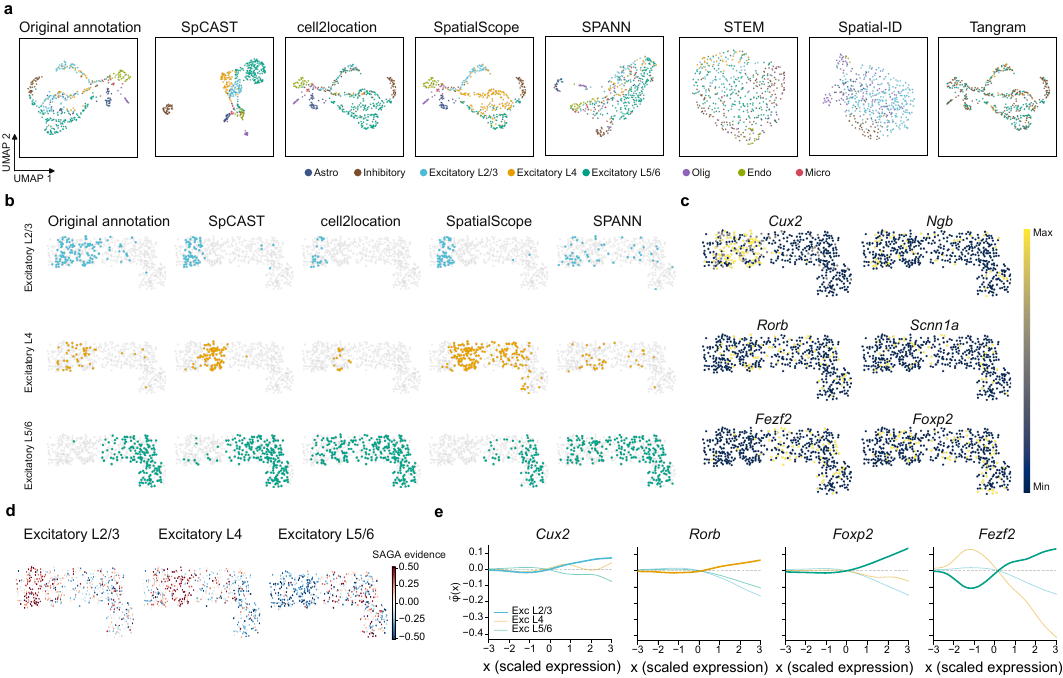}
    \caption{
    \textbf{SAGA links excitatory-neuron subtypes to laminar marker evidence in mouse visual cortex.}
    \textbf{a,} UMAP representations of the original annotations and predicted labels from SpCAST and six comparison methods.
    \textbf{b,} Spatial distributions of excitatory L2/3, L4 and L5/6 neurons in the original annotations and predictions from SpCAST, cell2location, SpatialScope and SPANN.
    \textbf{c,} Spatial expression of subtype-associated genes, including \textit{Cux2} and \textit{Ngb} for excitatory L2/3 neurons, \textit{Rorb} and \textit{Scnn1a} for excitatory L4 neurons, and \textit{Fezf2} and \textit{Foxp2} for excitatory L5/6 neurons.
    \textbf{d,} Spatial SAGA evidence maps for excitatory L2/3, L4 and L5/6 identities.
    \textbf{e,} Centered SAGA response functions for \textit{Cux2}, \textit{Rorb}, \textit{Foxp2} and \textit{Fezf2}.
    }
    \label{fig:visual_cortex_extended}
\end{extendedfigure}

\clearpage

% ============================================================
% Extended Data Figure 5
% ============================================================

\begin{extendedfigure}[p]
\centering
\includegraphics[
width=\textwidth,
% height=0.76\textheight,
keepaspectratio
]{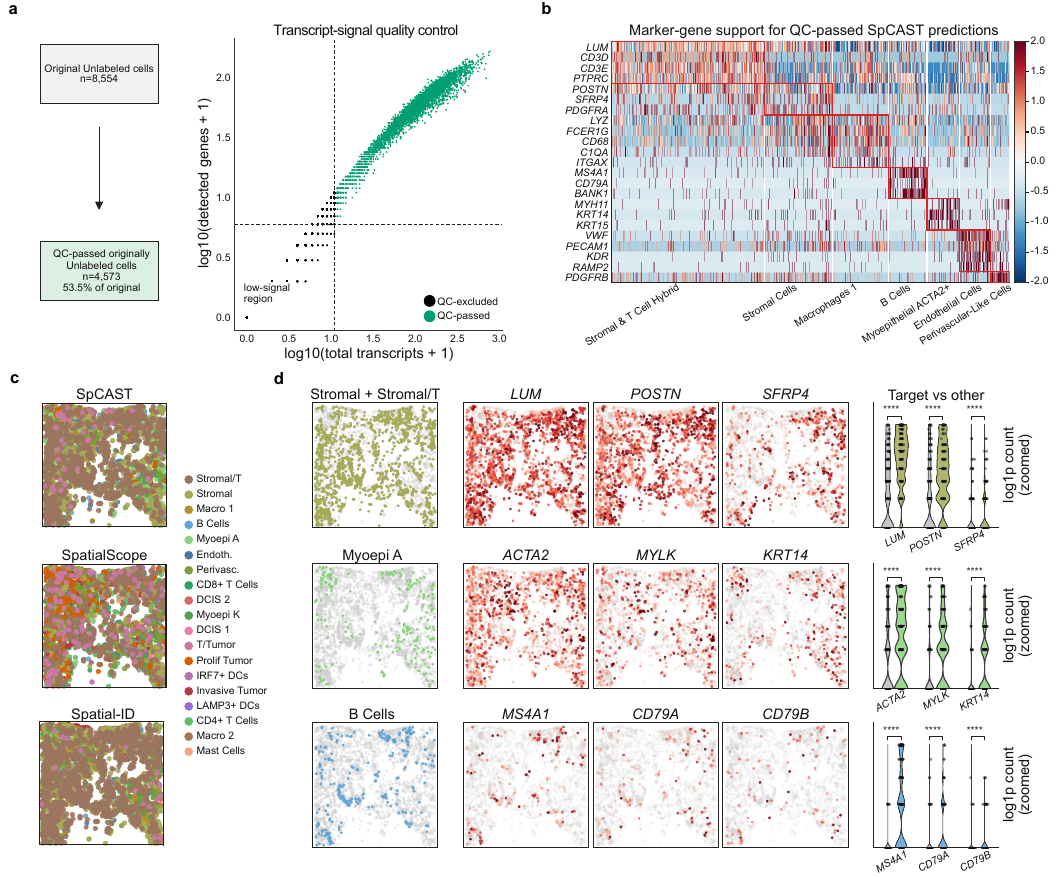}

\caption{
    \textbf{Marker-supported candidate identities for originally unlabelled cells in human breast cancer Xenium data.}
    \textbf{a,} Transcript-signal quality control of originally unlabelled cells. Cells with more than 10 total transcripts and more than 5 detected genes were retained, yielding 4,573 of 8,554 cells (53.5\%) for candidate-identity analysis. Total transcripts denote the summed panel counts per cell, and detected genes denote genes with nonzero counts.
    \textbf{b,} Marker-gene support heatmap for SpCAST candidate assignments among QC-passed originally unlabelled cells.
    \textbf{c,} Spatial candidate assignments from SpCAST, SpatialScope and Spatial-ID for QC-passed originally unlabelled cells.
    \textbf{d,} Representative SpCAST candidate assignments for stromal/stromal--T-cell-hybrid, myoepithelial ACTA2+ and B-cell populations, together with the spatial expression of corresponding marker genes and target-versus-other expression comparisons. Wilcoxon rank-sum tests; $^{****}P<0.0001$.
}

\label{fig:unlabeled_xenium_extended}

\end{extendedfigure}

\clearpage

% ============================================================
% Extended Data Figure 6
% ============================================================

\begin{extendedfigure}[p]
    \centering
    \includegraphics[
        width=\textwidth,
        % height=0.76\textheight,
        keepaspectratio
    ]{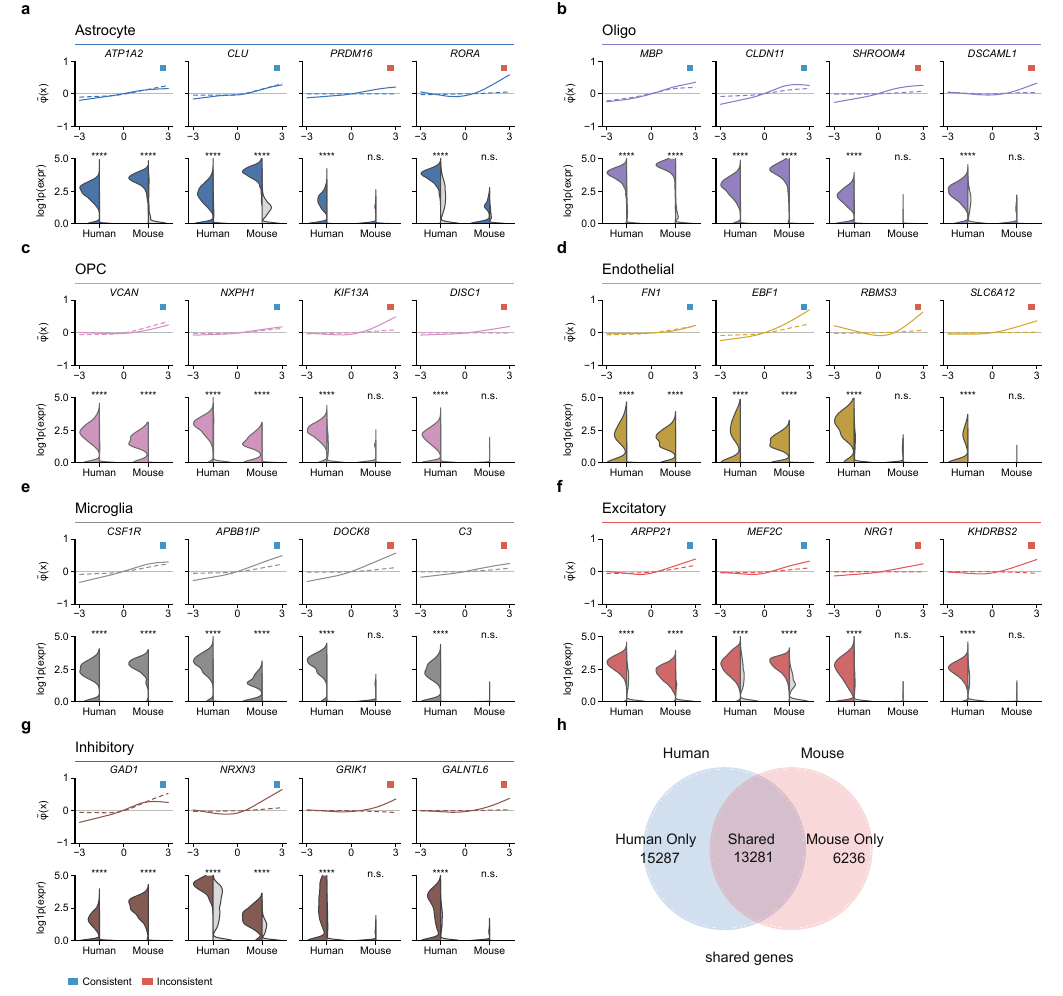}
    \caption{
    \textbf{Species-consistent and species-inconsistent gene responses across major brain cell types.}
    \textbf{a--g,} Representative genes for astrocytes (\textbf{a}), oligodendrocytes (\textbf{b}), OPCs (\textbf{c}), endothelial cells (\textbf{d}), microglia (\textbf{e}), excitatory neurons (\textbf{f}) and inhibitory neurons (\textbf{g}). For each gene, the upper plot shows centered SAGA response functions in the within-species and cross-species models, and the lower plot shows target-versus-other expression distributions in the human and mouse references. Each panel includes representative species-consistent and species-inconsistent genes. Target-versus-other expression differences were assessed using Wilcoxon rank-sum tests; $^{****}P<0.0001$; n.s., not significant.
    \textbf{h,} Overlap between the human Slide-tags query gene inventory and the ortholog-mapped mouse reference gene inventory, comprising 15{,}287 genes unique to the human query, 13{,}281 shared genes and 6{,}236 genes unique to the mouse reference.
    }
    \label{fig:cross_species_extended}
\end{extendedfigure}

\clearpage

\Heading{Supplementary Methods}\label{sec:supp_methods}

% ============================================================
% Supplementary Methods
% Purpose: provide implementation details that are not repeated
% in the main Methods.
% ============================================================

\heading{Slide-tags spatial expression enhancement}
\label{subsec:supp_slidetags_enhancement}

Because the Slide-tags query used in this study exhibited sparser single-nucleus expression profiles than the imaging-based spatial transcriptomics datasets, we implemented an optional local spatial-expression enhancement procedure for this dataset. The enhancement was performed after normalization and log$(1+x)$ transformation and before matching genes between the query and reference datasets.

For each spatial unit, $k=\min(30,N_{\mathrm{st}}-1)$ nearest spatial neighbours were first identified. The query expression matrix was projected by truncated singular value decomposition onto $p=\min(30,N_{\mathrm{st}},D_{\mathrm{st}})$ latent components, and neighbour similarity was quantified by the Pearson correlation $r_{ij}$ between the resulting latent representations. The $m=\min(4,k)$ neighbours with the highest correlations were retained.

Among the retained neighbours, only those with $r_{ij}>0$ were used, and their correlations were normalized to sum to one. Let $\mathcal{P}_i$ denote this set of positively correlated neighbours. The enhanced expression vector was
\begin{equation}
\widetilde{\mathbf{x}}_i
=
\begin{cases}
\displaystyle
\mathbf{x}_i+
\alpha
\sum_{j\in\mathcal{P}_i}
\frac{r_{ij}}
{\sum_{j'\in\mathcal{P}_i}r_{ij'}}
\mathbf{x}_j,
&
\mathcal{P}_i\neq\varnothing,\\[8pt]
\mathbf{x}_i,
&
\mathcal{P}_i=\varnothing.
\end{cases}
\label{eq:supp_slidetags_enhancement}
\end{equation}

The default enhancement coefficient in SpCAST is $\alpha=0.2$. For the head-to-head Slide-tags analyses reported in the main text, spatial-expression enhancement was disabled ($\alpha=0$) to avoid introducing an additional SpCAST-specific enhancement step that was not applied to the comparison methods. To assess parameter sensitivity, we additionally evaluated $\alpha\in{0.0,0.1,\ldots,0.9}$ in both the human-reference and mouse-reference settings of the same human Slide-tags query using random seeds 0, 42 and 2025 (Supplementary Table~\ref{tab:slidetags_enhancement_ablation}). The $\alpha=0$ setting corresponds to no spatial-expression enhancement. All annotation baselines received the original, unenhanced query expression matrix.

\heading{Additional KAN and MMD implementation details}
\label{subsec:supp_kan_mmd}

For the teacher encoder and query decoder, spline coefficients were initialized by least-squares fitting to small random target values drawn from $U(-0.02,0.02)$. For a layer with input dimension $d_{\mathrm{in}}$, base-branch scales were initialized from $U(-1/\sqrt{d_{\mathrm{in}}},1/\sqrt{d_{\mathrm{in}}})$ and spline-branch scales were initialized to $1/\sqrt{d_{\mathrm{in}}}$. The spline grid remained fixed during training. Under the KAN parameterization described in the main Methods, each edge contained eight spline coefficients.

The MMD loss was computed using the multiscale Gaussian kernels defined in the main Methods. Mean pairwise kernel values were evaluated for reference--reference, reference--query and query--query latent representations, and the resulting empirical MMD value was clipped at zero before being used as an optimization loss.

\heading{SAGA regularization and marker-score estimation}
\label{subsec:supp_saga_details}

Let $e_{igc}=\phi_{gc}(\widetilde{x}^{\mathrm{sc}}_{ig})$ denote the contribution of gene $g$ to class $c$ for reference cell $i$. The three SAGA regularization terms were
\begin{align}
\mathcal{L}_{\ell_1}
&=
\frac{1}{N_{\mathrm{sc}}DC}
\sum_{i,g,c}|e_{igc}|, \nonumber\\
\mathcal{L}_{\mathrm{sparse}}
&=
\sum_{g,c}
\sqrt{
\frac{1}{N_{\mathrm{sc}}}
\sum_i e_{igc}^{2}
+10^{-8}
}, \nonumber\\
\mathcal{L}_{\mathrm{smooth}}
&=
\sum_{g,c}
\sum_{k=1}^{K-2}
\left(
a_{gc,k+2}
-2a_{gc,k+1}
+a_{gc,k}
\right)^2 .
\label{eq:supp_saga_regularization}
\end{align}

Here, $D$ is the number of input genes, $C$ is the number of cell classes and $a_{gc,k}$ denotes the $k$th spline coefficient of the gene-$g$-to-class-$c$ function. SAGA used eight spline intervals with cubic B-splines, resulting in $K=11$ spline coefficients per edge.

Marker scores were calculated using teacher-predicted reference classes and the centred gene-level contributions defined in the main Methods. For each gene--class pair, $A_{gc}$ was the mean absolute centred contribution across reference cells, $D^{+}_{gc}$ was the positive difference in mean centred contribution between the target class and all remaining reference cells, and $U_{gc}$ was the AUROC obtained from the Mann--Whitney statistic comparing the two contribution distributions.

The final marker score was
\begin{equation}
\operatorname{SAGA}_{\mathrm{marker}}(g,c)
=
A_{gc}\,D^{+}_{gc}\,U_{gc}.
\label{eq:supp_saga_marker}
\end{equation}
Gene--class pairs with no target or no non-target reference cells were assigned a marker score of zero.

\heading{Benchmark implementations}
\label{subsec:supp_baselines}

Baseline methods were run using their official implementations where available. When a benchmark-specific implementation was required, the procedure followed the workflow described for the corresponding method. Method-specific preprocessing required by individual workflows was retained rather than imposing a single preprocessing procedure across all methods.

For the annotation benchmark, methods returning per-unit class abundances or probabilities were converted to discrete cell-type predictions after normalization where required. Tangram was run in cell-mapping mode, with the label of the highest-scoring mapped reference cell used as the primary discrete prediction. When class-level probability estimates were required, mapping scores were additionally aggregated by reference cell type.

For the reconstruction benchmark, all methods received the same artificially masked query within each benchmark instance, while method-specific preprocessing required by the corresponding workflows was retained. All reconstructed outputs were subsequently evaluated using the common procedure described below.

\heading{Controlled masking and reconstruction ablations}
\label{subsec:supp_reconstruction_details}

For the controlled masking benchmark, the artificially masked query was used as model input, and artificially masked entries were excluded from the query-reconstruction loss. The decoder was therefore optimized only against expression entries that remained visible during training.

Before evaluation, reconstructed matrices from all methods were aligned to the query gene order and subjected to the same per-gene scale calibration. Calibration parameters were estimated exclusively from visible, non-zero query entries. For gene $g$, let $\mu_g$ and $\sigma_g$ denote the mean and standard deviation of these observed values, and let $\widehat{\mu}_g$ and $\widehat{\sigma}_g$ denote the corresponding quantities for the predictions. For genes with at least five such entries and $\widehat{\sigma}_g\geq10^{-6}$, predictions were calibrated as
\begin{equation}
\widehat{x}^{\,\mathrm{cal}}_{ig}
=
\left[
\mu_g
+
\frac{\sigma_g}{\widehat{\sigma}_g}
\left(
\widehat{x}_{ig}-\widehat{\mu}_g
\right)
\right]_{+},
\qquad
[z]_{+}=\max(z,0).
\label{eq:supp_reconstruction_calibration}
\end{equation}

The calibration parameters were estimated only from visible entries but were subsequently applied to the complete predicted vector for the corresponding gene. Mean-ratio rescaling was used when $\widehat{\sigma}_g<10^{-6}$; genes with fewer than five visible non-zero entries were only clipped at zero. Artificially masked entries were not used to estimate calibration parameters.

In the SpCAST without sc-reference ablation, reference expression, reference labels, the reference classification loss and MMD alignment were removed. SpCAST-MLP retained the reference-guided objective but replaced the KAN encoder and decoder with dimension-matched feed-forward networks.

\heading{Computational scalability benchmark}
\label{subsec:supp_scalability}

The scalability benchmark was designed to evaluate computational performance as query size increased and was not used to assess annotation accuracy on synthetic data. Reference size, gene dimensionality and the number of classes were held fixed, whereas the number of query cells was progressively increased. All methods were evaluated on the same synthetic reference--query pairs.

Synthetic expression counts were generated using a fixed sampling procedure to maintain a consistent data distribution across query sizes. The query contained 313 genes; the fixed reference configuration and evaluated query sizes are described in the main Methods. Spatial coordinates were generated independently when required by individual methods.

Training time was measured from the start of model optimization to the completion of training. RAM and GPU memory usage were sampled every 2 s during model training by an independent monitoring process, and the maximum observed value during the training interval was reported.
\newpage
% Supplementary tables (Overleaf-ready standalone).
% Tables only — do not add a publisher SI cover page.
% Compile: pdflatex supplementary_tables.tex

% \documentclass[11pt,a4paper]{article}

% \usepackage[margin=18mm]{geometry}
% \usepackage{booktabs}
% \usepackage{multirow}
% \usepackage{pdflscape}
% \usepackage{graphicx}
% \usepackage{adjustbox}
% \usepackage{array}
% \usepackage{amsmath}
% \usepackage[T1]{fontenc}
% \usepackage{times}

\setcounter{table}{0}
\renewcommand{\tablename}{Supplementary Table}
\renewcommand{\thetable}{\arabic{table}}

% \begin{document}

% ------------------------------------------------------------
% Supplementary Table 1
% ------------------------------------------------------------
\refstepcounter{table}
\label{tab:paired_scrna_scst_datasets}
\noindent\textbf{Supplementary Table 1. Spatial transcriptomics datasets and their corresponding single-cell or single-nucleus transcriptomic reference datasets.}
Each row represents one reference--query pairing used in the analyses.
The human Slide-tags dataset was evaluated with both human and mouse
prefrontal-cortex references and is therefore listed twice.
% \par\smallskip
\begin{center}
\setlength{\tabcolsep}{3.2pt}
\renewcommand{\arraystretch}{1.12}
\small
\resizebox{\linewidth}{!}{%
\begin{tabular}{@{}llrrcllrrc@{}}
\toprule
\multicolumn{5}{c}{\textbf{Spatial transcriptomics query}} &
\multicolumn{5}{c}{\textbf{scRNA-seq reference}} \\
\cmidrule(lr){1-5}\cmidrule(lr){6-10}
\textbf{Description} &
\textbf{Technology} &
\textbf{Cells} &
\textbf{Genes} &
\textbf{Sections} &
\textbf{Description} &
\textbf{Technology} &
\textbf{Cells} &
\textbf{Genes} &
\textbf{Datasets} \\
\midrule
Mouse visual cortex & seqFISH+ & 497& 10,000 & 1 &
Mouse visual cortex & Smart-seq & 14,088& 34,041 & 1 \\

Mouse hypothalamus & MERFISH & 153,456 & 161 & 30 &
Mouse hypothalamus & 10X Chromium & 29,760 & 27,998 & 1 \\

Human non-small cell lung cancer & CosMx SMI & 87,606 & 980 & 20 &
Human non-small cell lung cancer & 10X Chromium & 51,896 & 33,694 & 1 \\

\multirow{2}{*}{Human prefrontal cortex} &
\multirow{2}{*}{Slide-tags} &
\multirow{2}{*}{4,065} &
\multirow{2}{*}{28,568} &
\multirow{2}{*}{1} &
Human prefrontal cortex & 10X Chromium & 153,473 & 26,747 & 1 \\

& & & & &
Mouse prefrontal cortex & 10X Chromium & 24,822 & 19,517 & 1 \\

Human breast cancer & Xenium & 167,780 & 313 & 1 &
Human breast cancer & scFFPE-seq & 27,472 & 37,143 & 1 \\
\bottomrule
\end{tabular}%
}
\end{center}
\par\bigskip

% ------------------------------------------------------------
% Supplementary Table 2
% ------------------------------------------------------------
\refstepcounter{table}
\label{tab:real_tech_summary}
\noindent\textbf{Supplementary Table 2. Technology-level cell-type transfer performance on real spatial transcriptomic datasets.}
Rows are methods and columns are scST technologies. Each technology is summarized by Accuracy (Acc.) and Weighted F1 (W-F1). For technologies with multiple datasets (MERFISH, CosMx SMI), values are means $\pm$ standard deviations across dataset-level seed averages; for single-dataset technologies (seqFISH+, Xenium, Slide-tags), values are means $\pm$ standard deviations across three random seeds (0, 42 and 2025). Bold indicates the highest mean within each technology and metric. Per-dataset results are provided in the accompanying Source Data file.
% \par\smallskip
\begin{center}
\setlength{\tabcolsep}{4.5pt}
\renewcommand{\arraystretch}{1.15}
\small
\resizebox{\linewidth}{!}{%
\begin{tabular}{@{}l*{5}{cc}@{}}
\toprule
\multirow{2}{*}{\textbf{Method}} & \multicolumn{2}{c}{\textbf{MERFISH}} & \multicolumn{2}{c}{\textbf{CosMx SMI}} & \multicolumn{2}{c}{\textbf{seqFISH+}} & \multicolumn{2}{c}{\textbf{Xenium}} & \multicolumn{2}{c}{\textbf{Slide-tags}} \\
\cmidrule(lr){2-3}\cmidrule(lr){4-5}\cmidrule(lr){6-7}\cmidrule(lr){8-9}\cmidrule(lr){10-11}
 & \textbf{Acc.} & \textbf{W-F1} & \textbf{Acc.} & \textbf{W-F1} & \textbf{Acc.} & \textbf{W-F1} & \textbf{Acc.} & \textbf{W-F1} & \textbf{Acc.} & \textbf{W-F1} \\
\midrule
SpCAST & \textbf{0.86 $\pm$ 0.03} & \textbf{0.86 $\pm$ 0.04} & 0.67 $\pm$ 0.16 & 0.64 $\pm$ 0.17 & \textbf{0.69 $\pm$ 0.00} & \textbf{0.68 $\pm$ 0.00} & \textbf{0.66 $\pm$ 0.00} & \textbf{0.66 $\pm$ 0.00} & \textbf{0.98 $\pm$ 0.00} & \textbf{0.98 $\pm$ 0.00} \\
cell2location & 0.74 $\pm$ 0.06 & 0.74 $\pm$ 0.07 & \textbf{0.72 $\pm$ 0.14} & \textbf{0.70 $\pm$ 0.14} & 0.63 $\pm$ 0.00 & 0.57 $\pm$ 0.00 & 0.28 $\pm$ 0.00 & 0.28 $\pm$ 0.00 & 0.97 $\pm$ 0.00 & 0.97 $\pm$ 0.00 \\
SPANN & 0.51 $\pm$ 0.03 & 0.51 $\pm$ 0.04 & 0.34 $\pm$ 0.07 & 0.41 $\pm$ 0.07 & 0.53 $\pm$ 0.00 & 0.51 $\pm$ 0.00 & 0.34 $\pm$ 0.00 & 0.34 $\pm$ 0.00 & 0.69 $\pm$ 0.00 & 0.70 $\pm$ 0.00 \\
Spatial-ID & 0.83 $\pm$ 0.08 & 0.82 $\pm$ 0.08 & 0.57 $\pm$ 0.19 & 0.53 $\pm$ 0.18 & 0.36 $\pm$ 0.00 & 0.29 $\pm$ 0.00 & 0.63 $\pm$ 0.00 & 0.63 $\pm$ 0.00 & 0.91 $\pm$ 0.00 & 0.93 $\pm$ 0.00 \\
SpatialScope & 0.83 $\pm$ 0.02 & 0.84 $\pm$ 0.02 & 0.61 $\pm$ 0.12 & 0.64 $\pm$ 0.13 & 0.62 $\pm$ 0.00 & 0.67 $\pm$ 0.00 & 0.64 $\pm$ 0.00 & 0.64 $\pm$ 0.00 & 0.94 $\pm$ 0.00 & 0.94 $\pm$ 0.00 \\
STEM & 0.62 $\pm$ 0.03 & 0.59 $\pm$ 0.03 & 0.17 $\pm$ 0.02 & 0.22 $\pm$ 0.02 & 0.48 $\pm$ 0.00 & 0.47 $\pm$ 0.00 & 0.39 $\pm$ 0.00 & 0.39 $\pm$ 0.00 & 0.75 $\pm$ 0.00 & 0.75 $\pm$ 0.00 \\
Tangram & 0.50 $\pm$ 0.10 & 0.47 $\pm$ 0.08 & 0.12 $\pm$ 0.01 & 0.14 $\pm$ 0.03 & 0.25 $\pm$ 0.00 & 0.19 $\pm$ 0.00 & 0.33 $\pm$ 0.00 & 0.34 $\pm$ 0.00 & 0.57 $\pm$ 0.00 & 0.57 $\pm$ 0.00 \\
\bottomrule
\end{tabular}%
}
\end{center}
\par\bigskip

\clearpage
% ------------------------------------------------------------
% Supplementary Table 3
% ------------------------------------------------------------
\refstepcounter{table}
\label{tab:marker_tech_summary}
\noindent\textbf{Supplementary Table 3. Technology-level reconstruction performance for reference-defined cell-type-associated genes.}
Rows are methods and columns are scST technologies. Each technology is summarized by Pearson (Pear.) and Spearman (Spear.) correlations. Values are means $\pm$ standard deviations across dataset--masking-rate combinations (masking rates 0.1, 0.3, 0.5 and 0.7), where each combination is first averaged over five random seeds (0, 7, 42, 123 and 2025).
% \par\smallskip

\begin{center}
\setlength{\tabcolsep}{4.5pt}
\renewcommand{\arraystretch}{1.15}
\small
\resizebox{\linewidth}{!}{%
\begin{tabular}{@{}l*{5}{cc}@{}}
\toprule
\multirow{2}{*}{\textbf{Method}}
& \multicolumn{2}{c}{\textbf{MERFISH}}
& \multicolumn{2}{c}{\textbf{CosMx SMI}}
& \multicolumn{2}{c}{\textbf{seqFISH+}}
& \multicolumn{2}{c}{\textbf{Xenium}}
& \multicolumn{2}{c}{\textbf{Slide-tags}} \\
\cmidrule(lr){2-3}
\cmidrule(lr){4-5}
\cmidrule(lr){6-7}
\cmidrule(lr){8-9}
\cmidrule(lr){10-11}
& \textbf{Pear.} & \textbf{Spear.}
& \textbf{Pear.} & \textbf{Spear.}
& \textbf{Pear.} & \textbf{Spear.}
& \textbf{Pear.} & \textbf{Spear.}
& \textbf{Pear.} & \textbf{Spear.} \\
\midrule

SpCAST
& \textbf{0.613 $\pm$ 0.070}
& \textbf{0.606 $\pm$ 0.046}
& \textbf{0.500 $\pm$ 0.073}
& \textbf{0.444 $\pm$ 0.075}
& -0.036 $\pm$ 0.005
& 0.061 $\pm$ 0.007
& \textbf{0.596 $\pm$ 0.040}
& \textbf{0.539 $\pm$ 0.038}
& \textbf{0.595 $\pm$ 0.009}
& 0.525 $\pm$ 0.006 \\

stPlus
& 0.459 $\pm$ 0.096
& 0.528 $\pm$ 0.054
& 0.366 $\pm$ 0.088
& 0.318 $\pm$ 0.068
& 0.190 $\pm$ 0.284
& 0.297 $\pm$ 0.014
& 0.543 $\pm$ 0.098
& 0.481 $\pm$ 0.052
& 0.565 $\pm$ 0.006
& 0.547 $\pm$ 0.001 \\

SpaGE
& 0.325 $\pm$ 0.084
& 0.507 $\pm$ 0.061
& 0.277 $\pm$ 0.062
& 0.268 $\pm$ 0.065
& 0.187 $\pm$ 0.270
& \textbf{0.311 $\pm$ 0.016}
& 0.333 $\pm$ 0.064
& 0.392 $\pm$ 0.051
& 0.574 $\pm$ 0.006
& \textbf{0.572 $\pm$ 0.004} \\

SpaIM
& 0.218 $\pm$ 0.085
& 0.199 $\pm$ 0.063
& 0.279 $\pm$ 0.034
& 0.250 $\pm$ 0.027
& 0.166 $\pm$ 0.004
& 0.143 $\pm$ 0.001
& 0.392 $\pm$ 0.040
& 0.396 $\pm$ 0.040
& 0.485 $\pm$ 0.012
& 0.489 $\pm$ 0.016 \\

SpatialScope
& 0.061 $\pm$ 0.022
& 0.164 $\pm$ 0.084
& 0.214 $\pm$ 0.048
& 0.319 $\pm$ 0.043
& 0.250 $\pm$ 0.005
& 0.159 $\pm$ 0.003
& 0.492 $\pm$ 0.017
& 0.363 $\pm$ 0.087
& 0.485 $\pm$ 0.001
& 0.550 $\pm$ 0.001 \\

Tangram
& 0.041 $\pm$ 0.012
& 0.149 $\pm$ 0.037
& 0.164 $\pm$ 0.073
& 0.209 $\pm$ 0.026
& \textbf{0.446 $\pm$ 0.015}
& 0.250 $\pm$ 0.002
& 0.434 $\pm$ 0.005
& 0.429 $\pm$ 0.008
& 0.384 $\pm$ 0.001
& 0.461 $\pm$ 0.000 \\

\bottomrule
\end{tabular}%
}
\end{center}
\par\bigskip

% ------------------------------------------------------------
% Supplementary Table 4
% ------------------------------------------------------------
\refstepcounter{table}
\label{tab:slidetags_enhancement_ablation}
\noindent\textbf{Supplementary Table 4. Sensitivity analysis of the spatial-expression enhancement coefficient for the human Slide-tags cortex query under two reference settings.}
Performance is reported as mean $\pm$ standard deviation across two reference settings and three random seeds ($n=6$ runs per $\alpha$). The default SpCAST enhancement coefficient is $\alpha=0.2$, whereas $\alpha=0$ corresponds to no spatial-expression enhancement. For the head-to-head Slide-tags comparisons reported in the main text, spatial-expression enhancement was disabled ($\alpha=0$) to avoid introducing an additional SpCAST-specific enhancement step that was not applied to the comparison methods.

% \par\smallskip

\begin{center}
\small
\begin{tabular}{@{}lcc@{}}
\toprule
Setting
& Accuracy
& Weighted F1 \\
\midrule

No enhancement ($\alpha=0.0$)
& $0.9711 \pm 0.0067$
& $0.9709 \pm 0.0066$ \\

$\alpha = 0.1$
& $0.9727 \pm 0.0062$
& $0.9724 \pm 0.0062$ \\

$\alpha = 0.2$
& $\mathbf{0.9736 \pm 0.0054}$
& $\mathbf{0.9733 \pm 0.0054}$ \\

$\alpha = 0.3$
& $0.9733 \pm 0.0055$
& $0.9731 \pm 0.0055$ \\

$\alpha = 0.4$
& $0.9725 \pm 0.0064$
& $0.9722 \pm 0.0065$ \\

$\alpha = 0.5$
& $0.9728 \pm 0.0066$
& $0.9724 \pm 0.0067$ \\

$\alpha = 0.6$
& $0.9706 \pm 0.0065$
& $0.9701 \pm 0.0067$ \\

$\alpha = 0.7$
& $0.9679 \pm 0.0092$
& $0.9675 \pm 0.0093$ \\

$\alpha = 0.8$
& $0.9673 \pm 0.0073$
& $0.9668 \pm 0.0074$ \\

$\alpha = 0.9$
& $0.9466 \pm 0.0383$
& $0.9435 \pm 0.0441$ \\

\bottomrule
\end{tabular}
\end{center}
\par\bigskip

% \end{document}

% ============================================================
% Supplementary Figures
% ============================================================

\clearpage
\section*{Supplementary Figures}

% Supplementary figure numbering: S1, S2, ...
\setcounter{figure}{0}
\renewcommand{\figurename}{Supplementary Figure}
\renewcommand{\thefigure}{S\arabic{figure}}

% ============================================================
% Supplementary Figure S1:
% KAN-versus-MLP architecture ablation
% ============================================================

\refstepcounter{figure}
\label{fig:supp_kan_mlp_ablation}

\begin{center}
    \vspace*{0.2em}

    \includegraphics[
        width=\textwidth,
        keepaspectratio,
        pagebox=artbox
    ]{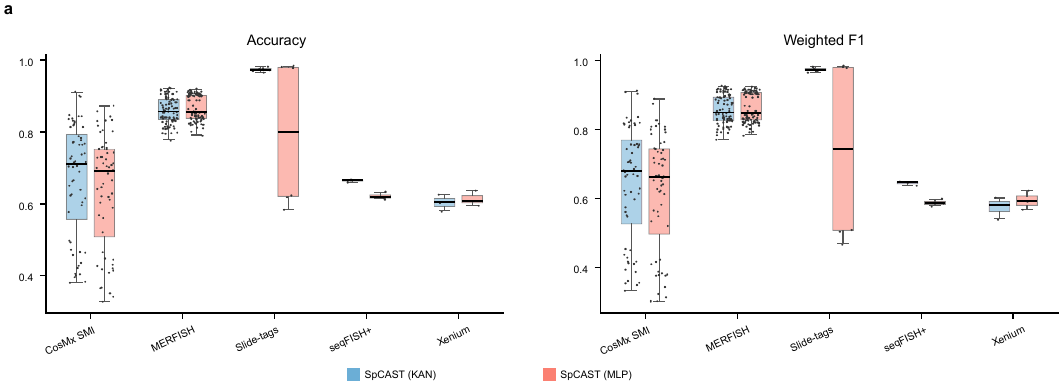}\par

    \vspace{0.8em}

    \noindent
    \parbox{0.96\textwidth}{%
        \small
        \textbf{\figurename~\thefigure.}
        \textbf{Comparison of KAN- and MLP-based SpCAST variants across spatial transcriptomics technologies.}
        \textbf{a,}
        Accuracy (left) and weighted F1 (right) are compared between the
        complete SpCAST model with a KAN encoder--decoder and a
        dimension-matched SpCAST (MLP) ablation across CosMx SMI,
        MERFISH, Slide-tags, seqFISH+ and Xenium datasets.
        Boxplots summarize the distributions of the benchmark observations
        shown as individual points.
    }
\end{center}

\clearpage

% ============================================================
% Supplementary Figure S2:
% Additional SAGA marker maps
% ============================================================

\refstepcounter{figure}
\label{fig:supp_saga_additional_marker_maps}

\begin{center}
    \vspace*{0.2em}

    \includegraphics[
        width=\textwidth,
        keepaspectratio,
        pagebox=artbox
    ]{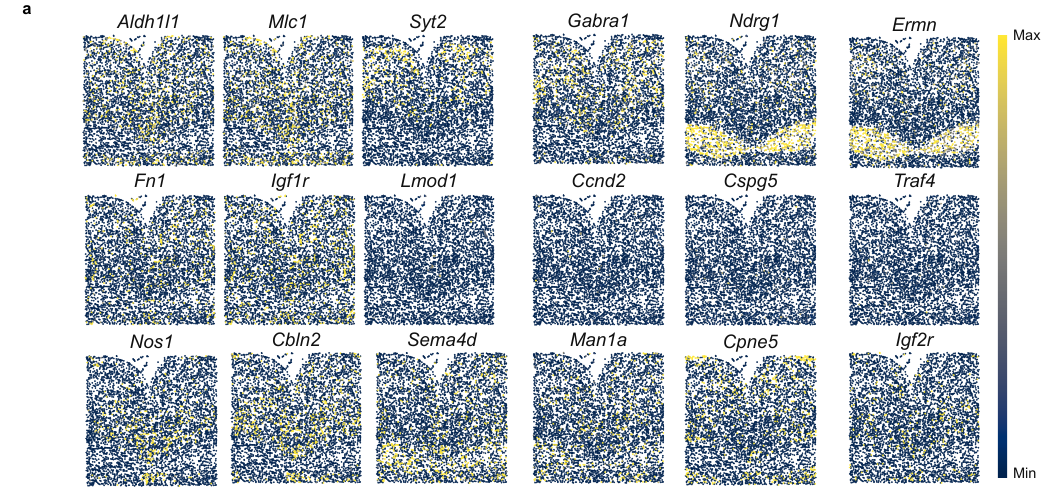}\par

    \vspace{0.8em}

    \noindent
    \parbox{0.96\textwidth}{%
        \small
        \textbf{\figurename~\thefigure.}
        \textbf{Additional spatial expression maps of SAGA-prioritized marker-gene candidates in mouse hypothalamus MERFISH data.}
        \textbf{a,}
        Spatial expression patterns are shown for
        \textit{Aldh1l1} and \textit{Mlc1} (astrocytes);
        \textit{Syt2} and \textit{Gabra1} (inhibitory neurons);
        \textit{Ndrg1} and \textit{Ermn} (mature oligodendrocytes);
        \textit{Fn1} and \textit{Igf1r} (endothelial cells);
        \textit{Lmod1} and \textit{Ccnd2} (mural cells);
        \textit{Cspg5} and \textit{Traf4} (immature oligodendrocytes);
        \textit{Nos1} and \textit{Cbln2} (excitatory neurons);
        \textit{Sema4d} and \textit{Man1a} (microglia);
        and \textit{Cpne5} and \textit{Igf2r} (ependymal cells).
        The colour scale indicates relative expression from minimum to
        maximum.
    }
\end{center}

\clearpage

% ============================================================
% Supplementary Figure S3:
% Xenium marker-expression heatmap
% ============================================================

\refstepcounter{figure}
\label{fig:supp_xenium_marker_heatmap}

\begin{center}
    \vspace*{0.2em}

    \includegraphics[
        width=\textwidth,
        keepaspectratio,
        pagebox=artbox
    ]{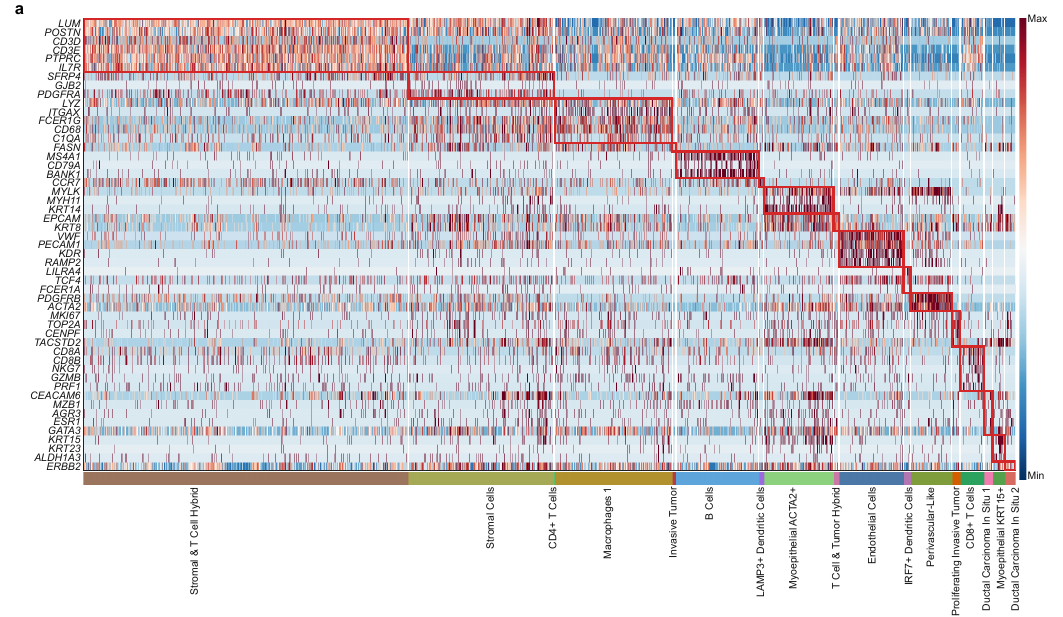}\par

    \vspace{0.8em}

    \noindent
    \parbox{0.96\textwidth}{%
        \small
        \textbf{\figurename~\thefigure.}
        \textbf{Marker-gene expression across annotated cell classes in human breast cancer Xenium data.}
        \textbf{a,}
        Heatmap of representative marker genes across cells grouped by their
        original non-Unlabeled annotations.
        Rows correspond to genes and columns to individual cells ordered by
        cell class.
        The lower annotation bar denotes cell-class membership, red outlines
        mark class-associated expression blocks and the colour scale indicates
        relative expression from minimum to maximum.
    }
\end{center}

\clearpage

% ============================================================
% Supplementary Figure S4:
% Xenium marker-based validation
% ============================================================

\refstepcounter{figure}
\label{fig:supp_xenium_marker_validation}

\begin{center}
    \vspace*{0.2em}

    \includegraphics[
        width=\textwidth,
        keepaspectratio,
        pagebox=artbox
    ]{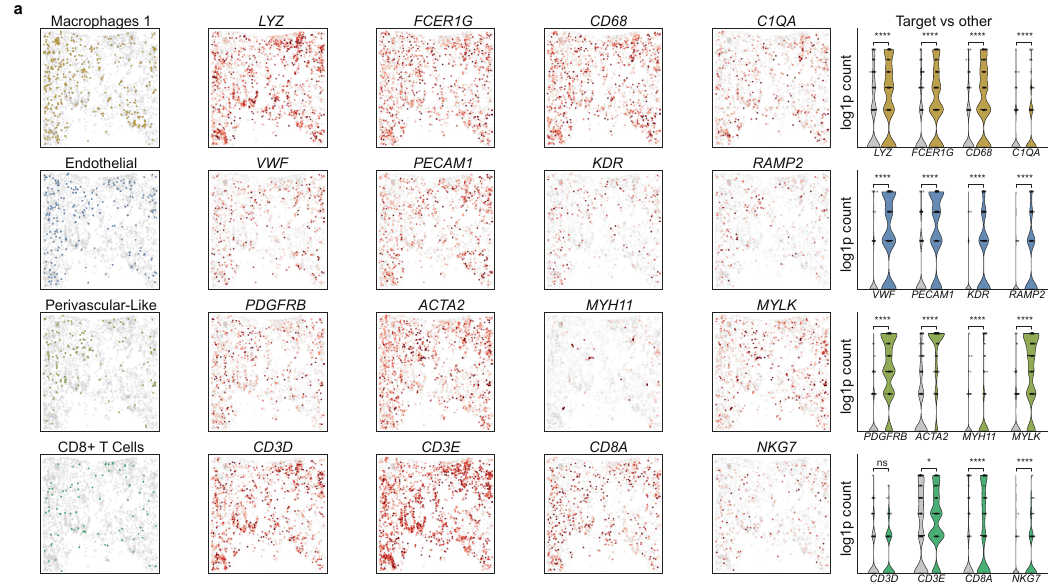}\par

    \vspace{0.8em}

    \noindent
    \parbox{0.96\textwidth}{%
        \small
        \textbf{\figurename~\thefigure.}
        \textbf{Marker-based validation of additional SpCAST-assigned cell types among originally Unlabeled cells in human breast cancer Xenium data.}
        \textbf{a,}
        Rows show SpCAST-assigned Macrophages 1, Endothelial,
        Perivascular-Like and CD8+ T-cell candidates.
        The first column shows the spatial distribution of each assigned
        population; the middle columns show the corresponding marker-gene
        expression maps
        (\textit{LYZ}, \textit{FCER1G}, \textit{CD68} and \textit{C1QA};
        \textit{VWF}, \textit{PECAM1}, \textit{KDR} and \textit{RAMP2};
        \textit{PDGFRB}, \textit{ACTA2}, \textit{MYH11} and \textit{MYLK};
        and \textit{CD3D}, \textit{CD3E}, \textit{CD8A} and \textit{NKG7},
        respectively).
        Violin plots compare log1p-transformed marker counts between the
        target population and all other cells.
        Target-versus-other expression differences were assessed using
        Wilcoxon rank-sum tests; $^{*}P < 0.05$,
        $^{****}P < 0.0001$; n.s., not significant.
    }
\end{center}

\clearpage

% ============================================================
% End of Supplementary Figures
% ============================================================

\end{document}